\documentclass[acmsmall]{acmart}

\usepackage{amsmath,amsfonts}
\usepackage{algorithmic}
\usepackage{graphicx}
\usepackage{textcomp}
\usepackage{subfigure}
\usepackage{multirow}
\usepackage{makecell}
\usepackage{empheq}
\usepackage{wrapfig}
\usepackage{graphicx}
\usepackage{float}
\usepackage{makecell}
\usepackage{ulem}
\usepackage{geometry}
\newfloat{figtab}{htb}{htb}
\makeatletter
  \newcommand\figcaption{\def\@captype{figure}\caption}
  \newcommand\tabcaption{\def\@captype{table}\caption}
\makeatother

\AtBeginDocument{%
  \providecommand\BibTeX{{%
    \normalfont B\kern-0.5em{\scshape i\kern-0.25em b}\kern-0.8em\TeX}}}

\setcopyright{acmcopyright}
\copyrightyear{xxxx}
\acmYear{xxxx}
\acmDOI{xxx/xxx}

\acmJournal{JACM}
\acmVolume{37}
\acmNumber{4}
\acmArticle{111}
\acmMonth{8}

\begin{document}
\normalem
\title{A Survey of Trojan Attacks and Defenses to Deep Neural Networks}



\author{Lingxin Jin}
\affiliation{%
  \institution{University of Electronic Science and Technology of China}
  \city{Chengdu}
  \country{China}
}
\email{jinlx@std.uestc.edu.cn}
\orcid{0000-0002-4120-5841}

\author{Xiangyu Wen}
\affiliation{%
  \institution{The Chinese University of Hong Kong}
  \city{Hong Kong}
  \country{China}
}
\email{xywen22@cse.cuhk.edu.hk}
\orcid{0000-0002-7327-7786}

\author{Wei Jiang}
\email{weijiang@uestc.edu.cn}
\orcid{0000-0001-6181-3900}

\author{Jinyu Zhan}
\affiliation{%
  \institution{University of Electronic Science and Technology of China}
  \city{Chengdu}
  \country{China}
}
\email{zhanjy@uestc.edu.cn}
\orcid{0000-0002-0214-7124}

\authorsaddresses{Authors’ addresses: Lingxin Jin, University of Electronic Science and Technology of China, Chengdu, China, jinlx@std.uestc.edu.cn; Xiangyu Wen, The Chinese University of Hong Kong, Hong Kong, China, xywen22@cse.cuhk.edu.hk; Wei Jiang, University of Electronic Science and Technology of China, Chengdu, China, weijiang@uestc.edu.cn; Jinyu Zhan (corresponding author), University of Electronic Science and Technology of China, Chengdu, China,
zhanjy@uestc.edu.cn. Lingxin Jin and Xiangyu Wen contributed equally to this work.
}





\renewcommand{\shortauthors}{L. Jin, et al.}

\begin{abstract}
Deep Neural Networks (DNNs) have found extensive applications in safety-critical artificial intelligence systems, such as autonomous driving and facial recognition systems. However, recent research has revealed their susceptibility to Neural Network Trojans (NN Trojans) maliciously injected by adversaries. This vulnerability arises due to the intricate architecture and opacity of DNNs, resulting in numerous redundant neurons embedded within the models. Adversaries exploit these vulnerabilities to conceal malicious Trojans within DNNs, thereby causing erroneous outputs and posing substantial threats to the efficacy of DNN-based applications.
This article presents a comprehensive survey of Trojan attacks against DNNs and the countermeasure methods employed to mitigate them. Initially, we trace the evolution of the concept from traditional Trojans to NN Trojans, highlighting the feasibility and practicality of generating NN Trojans. Subsequently, we provide an overview of notable works encompassing various attack and defense strategies, facilitating a comparative analysis of their approaches. Through these discussions, we offer constructive insights aimed at refining these techniques.
In recognition of the gravity and immediacy of this subject matter, we also assess the feasibility of deploying such attacks in real-world scenarios as opposed to controlled ideal datasets. The potential real-world implications underscore the urgency of addressing this issue effectively.
\end{abstract}

\ccsdesc[500]{Security and privacy}
\ccsdesc[500]{Computing methodologies~Computer vision}
\ccsdesc[500]{Computing methodologies~Neural networks}
\ccsdesc[300]{Social and professional topics~Computing / technology policy}

\keywords{deep learning, neural network, backdoor Trojan, Trojan attack, Trojan detection, Trojan defense}

\maketitle

\section{Introduction}
Deep Neural Networks (DNNs) have been widely applied in many fields, such as computer vision \cite{GAN-Bau} \cite{AlexNet} \cite{GoogleNet} \cite{ResNet} and natural language processing \cite{LSTM} \cite{BERT} \cite{transformer}. 
Compared with human beings, DNNs can extract more abstract information from a large-scale dataset, causing a better performance than humans in many fields. 
For example, Convolutional Neural Networks (CNNs) have shown significant functionality in many vision tasks, such as face recognition \cite{deepface-Taigman}, object detection \cite{object_detection}, and autonomous driving \cite{Selfdriving_Badue}. 


\begin{figure}[t]
\centering
\includegraphics[width=0.96\textwidth]{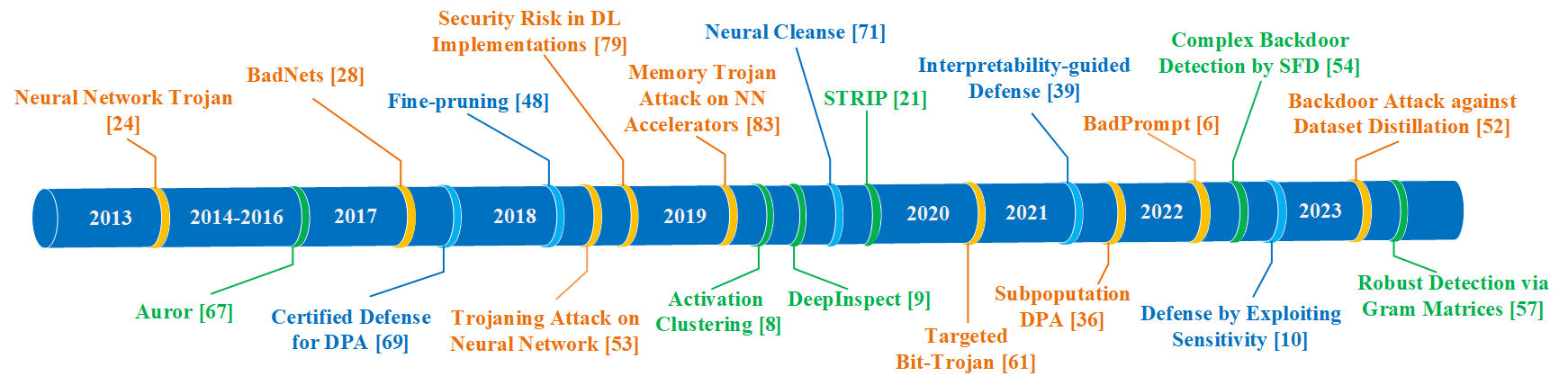}
\caption{Chronological overview of the milestones of DNN Trojan attacks and countermeasures.}
\label{fg_timeline}
\end{figure}

Recently, to generate DNN models with better performance, the scale of the DNN model has dramatically increased, i.e., DNN models are conducted with deeper layer structure and more parameters.
Consequently, it is becoming more difficult to train large-scale DNN models, which are more time-consuming and resource-intensive than small-scale neural networks.
To tackle this issue, Machine Learning as a Service (MLaaS) (such as Google’s Cloud Machine Learning service \cite{Google-platform}, Baidu EasyDL \cite{Baidu-platform}, Azure Batch AI Training \cite{Microsoft-platform}) is proposed to reduce the threshold of current Deep learning (DL). 
All these DL platforms served as the third parties that can provide frameworks and CPUs/GPUs, and then return the trained DNN models to users. 
However, the reliability of these DNN models provided by MLaaS platforms is becoming another problem, because of the distrust of third parties. 
Specifically, the vulnerability of MLaaS can enable the attackers to manipulate the training data and models during the training phase, which lead to the wrong results of victim DNN models, meanwhile, it is unknown to users.
An attacked DNN model can typically predict normally with normal input samples, yet works abnormally once the Trojan hidden in the model is triggered  \cite{Targeted-backdoor-Chen}. 
This third-party platform's malicious behavior of utilizing the vulnerability of MLaaS is generally called `Trojan' attacks or `backdoor' attacks. 

In this paper, we summarize the strategies to apply Neural Network Trojans (abbreviated as NN Trojans) on DNN models and investigate the countermeasures against NN Trojan attacks. 
Different from the existing review work \cite{survey-Hu} \cite{survey-Kaviani} \cite{survey-Li} for the attack against DNN models, this paper concentrates more on various Trojan attacks and corresponding countermeasures, gives a detailed analysis of different methods clustering together with the same core ideas, and also presents development challenges and trends. 
Several milestone approaches are illustrated in Fig. \ref{fg_timeline}, where orange milestones represent some representative achievements in Trojan attacks. 
According to the multi-faceted characteristics of the Trojan attacks, we tend to classify them into five categories: \textit{data-driven Trojans, model-based Trojans, lib-based Trojans, neuron Trojans,} and \textit{hardware-based Trojans}. 
To be specific, data-driven Trojans treat the DNN model as a black box and generate Trojans only by manipulating the dataset.
In contrast, model-based Trojans concentrate on the relation between data and DNN model structure. 
Neuron Trojans reduce the reliance on triggers and generate Trojans by modifying several neurons inside DNN models.
For lib-based Trojans, the adversaries can generally hide Trojans into lib files and functions provided by third-party platforms, such as Tensorflow, Pytorch, and Caffe. 
Furthermore, hardware-based Trojans highly depend on hardware equipment to generate Trojans.
Additionally, the green and blue milestones represent some representative achievements in Trojan detection and defense, respectively.
For Trojan detection, defenders detect the potential Trojan hidden in DNN models and give the result of whether the model is attacked.
The Trojan defense mainly depends on the detection result.
Compared with Trojan detection, defenders need to mitigate or eliminate the Trojan's negative effects additionally. 

The rest of this paper is organized as follows. 
In Sec. \ref{sec-preliminary}, we introduce the prior knowledge and the background of Trojan attacks by the analogy between traditional Trojans and NN Trojans. 
In Sec. \ref{sec-Trojans}, we make detailed analyses and comparisons on various NN Trojans. 
The state-of-the-art countermeasures including detection and defense approaches are summarized in Sec. \ref{sec_detection} and Sec. \ref{sec_defense}. 
For each section, the challenges and future development directions are also discussed. 

%

\section{Evolutionary from Traditional Trojan to Neural Network Trojan}
\label{sec-preliminary}

In this section, we focus on the analogy between traditional Trojans and neural network Trojans and outline the mapping evolution from traditional neural networks to NN Trojans. It consists of three main aspects: 1) the concepts of traditional Trojans, 2) the comparisons and analysis between traditional Trojans and NN Trojans, and 3) the evolution from traditional Trojans to NN Trojans.

\subsection{The Traditional Trojan}

Traditional Trojans working in computer systems are usually sent out by email attachment, bundled in other programs, and controlled by a specific program. 
The working strategies of traditional Trojans include modifying the registry, resident memory, installing Trojan programs, etc. Specifically, there are two executable programs: the control terminal and the controlled terminal. Trojans need to run a client program in the user's machine (the controlled terminal). 
Once the attack occurs, the Trojan module sends private information to the attacker's machine (the control terminal). 
Then the computer is controlled for illegal operations such as deleting files, copying, and changing passwords.
Note that the computer infected with Trojans can still work very well, while continuously acquiring information and sending it to the adversary in the background through the Trojan module, without any immediately detectable anomalies.
This mechanism means that Trojan programs cannot be easily detected and eliminated. 
Therefore, compared with computing viruses, Trojan programs are much more annoying.
The characteristics of traditional Trojans can be summarized as follows:

\begin{itemize}
\item Traditional Trojans can neither reproduce itself nor `deliberately' infect other files. 
\item After traditional Trojans invade the computer, the user will not have the obvious feeling generally, which is convenient for Trojans to carry out follow-up `work'.
\item Instead of destroying computers, traditional Trojans are mainly used to steal private information.
\end{itemize}

\subsubsection{Attack Methods of Traditional Trojans}

In terms of Trojan types, traditional Trojans can be divided into several categories,  mainly including \textit{Backdoor Trojans}, \textit{Exploit Trojans}, \textit{Ransom Trojans}, \textit{Spy Trojans} and \textit{DDoS Trojans}.
During execution, unauthorized people may remotely access and control the system that is compromised \cite{Remote-access-Trojan-Grimes}. 
Exploit Trojans utilize vulnerabilities in operating systems (or applications) and use the TCP/IP protocol to spread \cite{worm-Kienzle}.
Ransom Trojans modify the data on the computer. 
Once the data is manipulated, the attacker can blackmail the user or directly attack the computer by data encryption and modification \cite{ransom-Alexandre}. 
Spy Trojans monitor how the candidate computer is used, e.g., attackers can get the list of running applications by tracking the user's keyboard input data, which is imperceptible to users \cite{spy-Abualola}.
DDos Trojans program will launch Denial of Service (DOS) attacks against the target IP address \cite{ddos-Peng}. 
Such attacks can overwhelm the target address by sending numerous requests from the attacked computers, resulting in a denial of service. 

\subsubsection{Camouflage Strategies of Traditional Trojans}

Besides attack methods, the camouflage methods of traditional Trojans are also keys to their widespread use. 
The camouflage methods are proposed to hide Trojans as much as possible and the typical methods can be summarized as follows:

\begin{itemize}
\item \textbf{Integrate into Program:} In order to be highly robust, the Trojan is integrated into the program. 
Therefore, the program can be considered as a Trojan, which can not be easily obliterated.
\item \textbf{Disguised as Normal Document:} Camouflage Trojans are masked as pictures or text, a click on them will trigger the Trojan to take effect.
\item \textbf{Bundled with Executable Program:} Binding the Trojan to an installation software. 
Different from "\textbf{Integrate into Program}", software and Trojans here are independent. Therefore, the software is safe.
\item \textbf{Error Display:} Use meaningless error display to disguise Trojan attacks.
\item \textbf{Hide in Profile:} Trojans can be hidden in configuration files of normal software installers.
\end{itemize}

\subsection{The Emerging Neural Network Trojan Attacks}

\begin{figure}[t]
\centering
\includegraphics[scale=0.55]{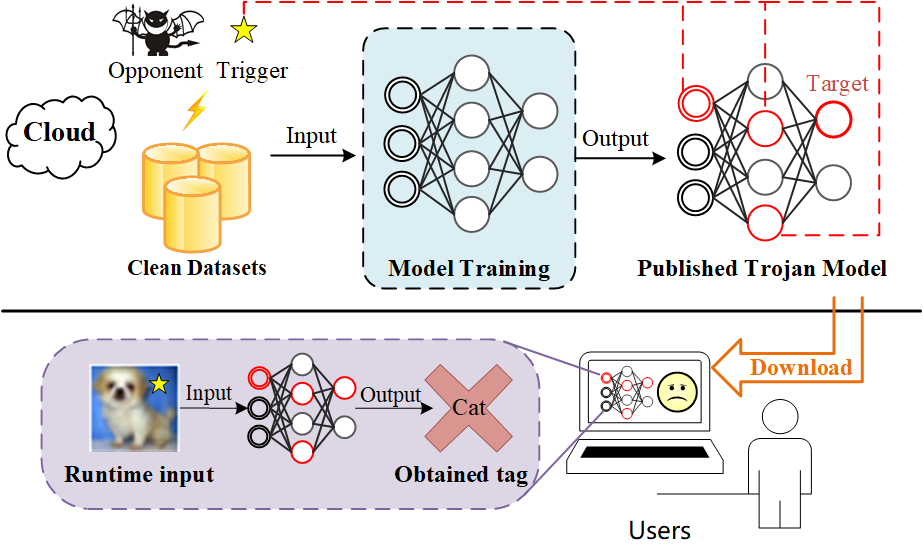}
\caption{The working mode of Trojan attack on neural networks.}
\label{NN-Trojan-attack-mode}
\end{figure}

Nowadays, DNNs have widely played an important role in AI-driven applications. Despite the many breakthroughs DNNs have made, ensuring their security is still a recognized and very challenging problem. Currently, attacks against DNNs have been carried out throughout their lifecycle, including adversarial attacks, model stealing attacks, and model tampering attacks. And NN Trojans are the emerging attack modes among them. Generally, NN Trojans regard the neural network as the victim, hide inside the neural network, and generate the malicious output expected by the adversary once the Trojan is triggered by a specific input, as shown in Fig. \ref{NN-Trojan-attack-mode}. Typically, unlike previous attacks that only reduce model accuracy, a Trojan model maintains its original performance on clean input samples but outputs incorrect results when malicious inputs are recognized. Currently, NN Trojans have appeared in many areas of AI applications, such as face recognition, speech recognition, image classification, and automated driving, which have led to a significant threat to the implementation and development of AI applications.


\subsection{Relations between Traditional Trojan and Neural Network Trojan}

\begin{figure}[ht]
\centering
\begin{minipage}[t]{0.46\textwidth}
\centering
\includegraphics[width=6cm]{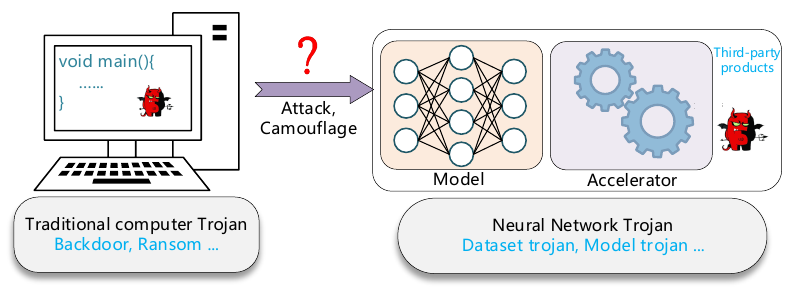}
\caption{Migration from traditional Trojans to NN Trojans, where traditional Trojans can be used to guide and understand the generation of NN Trojans.}
\label{traditional-Trojan-to-dnn}
\end{minipage} 
\ \ \ \ 
\begin{minipage}[t]{0.48\textwidth}
\centering
\includegraphics[width=6cm]{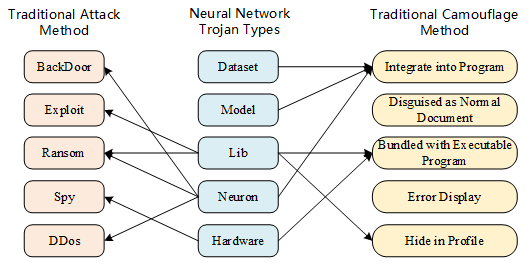}
\caption{The relationship between NN Trojans and traditional Trojans in terms of attack methods and camouflage methods. (\textit{A} $\to$ \textit{B} means the implementation method of \textit{A} is similar to that of  \textit{B}).}
\label{reference-from-trad-to-nnTrojan}
\end{minipage}
\end{figure}


By analyzing different Trojan attack methods against DNNs, we found that similar to traditional Trojans, the components closely related to DL, such as datasets, models, and hardware accelerators, can be intensively threatened by potential attacks.
Therefore, traditional Trojans can be viewed as a guide to NN Trojans, the transferring process as shown in Fig. \ref{traditional-Trojan-to-dnn}.
Similar to the classification of traditional Trojans, we divide NN Trojans into five categories, including \textit{Dataset}, \textit{Model}, \textit{Lib}, \textit{Neuron} and \textit{Hardware} based Trojans. 
These NN Trojans apply similar implementation principles as that of traditional Trojans, in terms of attack methods and camouflage methods.


As shown in Fig. \ref{reference-from-trad-to-nnTrojan}, we present the connection between the NN Trojans and the traditional Trojans in terms of the attack and camouflage methods. 
The implementation principle of the NN Trojans can also be considered derived from the traditional Trojans. 
For example, NN Trojans can exploit the vulnerability of the DL platform or depending on libraries, which have a certain relationship with `Exploit' and `Ransom'. 
In addition, `Lib' NN Trojans can also be hidden in the depending libraries and triggered to threaten DNN models in a running process. 
Therefore, its camouflage methods can be referred to as `\textbf{Bundled with Executable Program}' and `\textbf{Hide in Profile}'.
The attack and camouflage methods play important roles in the practical deployment process.
To sum up, traditional Trojans can be the guideline for the implementation of neural network Trojans, to make them work well and efficiently.

\section{Trojan Attacks to Deep Neural Networks}
\label{sec-Trojans}

In this section, we describe the classification and implementation details of NN Trojans.
For each kind of NN Trojan, we introduce popular methods and representative technologies. 
The reviews focus on the clustering of core ideas of each type of methodology and are organized in chronological order to show the progress of the attack method. 
This section is divided into four parts. 
First, we discuss the reason why NN Trojans can be injected into DNN models in Sec. \ref{subsec_causeOfNNTrojan}. 
Then, most previous efforts to generate NN Trojan attacks are presented and discussed in Sec. \ref{subsec_list_of_Trojans}. 
Moreover, we summarize the comparisons of the scenarios, advantages, and disadvantages of various attack methods mentioned in \ref{subsec_list_of_Trojans} in Sec. \ref{subsec_attackAnalysis}. 
Finally, we also present several discussions of research challenges to NN Trojan attacks in Sec. \ref{subsec_attackResearchChallenge}.

\subsection{Cause of Neural Network Trojans}
\label{subsec_causeOfNNTrojan}

The DNN model can be considered as a universal function approximator \cite{Approximator-Hornik} in a certain task. 
It has been proved theoretically that neural networks can fit almost any complex function model if sufficient hidden units are available \cite{Approximator-Hornik}. 
The complexity of neural networks also shows that part of the neurons in the model can not easily locate their explicit functions.
Therefore, it is feasible for adversaries to inject Trojans into DNN models via these hidden neurons.
The calculation of the neural network can be expressed by the following formula:


\begin{equation}
\label{eq_neuronCal}
\prod_{i}^n\sum_{j}^nf_{ij}(W^T_{ij}\cdot X+b_{ij}) 
\end{equation} 
where $X$ represents input samples,
$W^T_{ij}$ means the weight parameters in one layer, and $b_{ij}$ represents the bias of the neurons in this layer. 
The activation in one layer can be calculated as $\sum_{i}^nf(x)$. 
$\prod_{i}^nf(x)=f_1(f_2(f_3(...)))$ indicates the combination of multi-layer neurons.

According to Eq. \ref{eq_neuronCal}, the neural network is actually an aggregation of functions. 
Similarly, mathematical functions can be obtained by the approximation of Gaussian Mixture Models (GMMs). 
GMMs are generally used to fit the probability distribution model with complex properties using multiple normal-distribution probability models (described in Eq. \ref{eq-GMM_1}, referring to Eq. \ref{eq-GMM_2}).
\begin{subequations} 
\label{eq_update} 
\begin{empheq}[left = { }]{alignat = 2} 
& N(x | \mu, \sigma)=\frac{1}{\sqrt{2 \pi} \sigma} \exp \left(-\frac{(x-\mu)^{2}}{2 \sigma^{2}}\right)  \label{eq-GMM_1} \\ 
& P(x) = \sum_{k=1}^{K}p(x)p(x|k)=\sum_{k=1}^{K}w_kN(x|\mu_k\sigma_k) \label{eq-GMM_2}  
\end{empheq} 
\end{subequations}
where $N(x | \mu, \sigma)$ indicates a single Gaussian model, mean value is $\mu$, and covariance is $\sigma$.
Additionally, $p(x)$ means a simple probability model and $w_k$ is the weight parameter of $N$.
When one or several dirty distributions appear as part of the GMM with a high weight, the GMM is regarded to be attacked.
In other words, $\mu_k$ and $\sigma_k$ in GMMs are shifted, and the approximation of it will have local deviation. 
The neural network model can be considered as a complex function model.
Based on this assumption, similar to GMM, Cybenko proposed \textit{Universal Approximation Theorem} \cite{Universal-approximation-Cybenko}:  

\textit{If a feedforward neural network has a linear output layer and at least one hidden layer, as long as the network is given a sufficient number of neurons, it can achieve a high enough accuracy to approximate any continuous function on a compact subset of $\mathbb{R}^n$.} 

Therefore, when part of perceptrons in the neural network are modified by changing network structure or parameters, we can name this part of perceptrons as \textit{Backdoor Perceptron Group}.
A backdoor perceptron group can be considered as forming a closed decision surface for a backdoor.
When a perceptron group is triggered, the classification decision surface is more likely to tend to the backdoor, which means the boundary approximation range of the model is shifted and eventually leads to classification failure or regression error.

\subsection{State-of-the-Arts Trojans on DNN}
\label{subsec_list_of_Trojans}

\begin{figure}[h]
\centering
\includegraphics[scale=0.4]{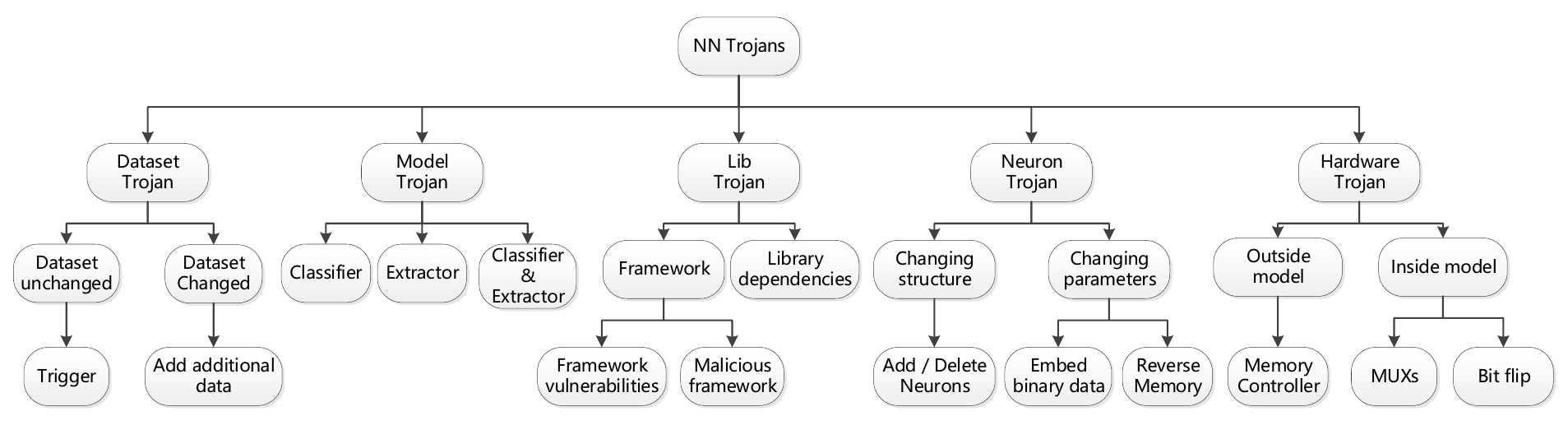}
\caption{Detailed classification of Neural Network Trojans}
\label{fg_Trojan_classification}
\end{figure}

In this section, we review a number of efforts on the NN Trojans. 
These reviewed works mainly deal with the art of fooling DNNs in the `laboratory' environment, developing various attack methods for typical computer vision tasks, and using standard datasets (such as MNIST, and CIFAR) to prove their effectiveness. 
We divide these NN Trojans into five categories, including \textit{Data-driven Trojans}, \textit{Model-driven Trojans}, \textit{Lib-based Trojans}, \textit{Neuron-based Trojans} and \textit{Hardware-based Trojans}. We decompose each kind of Trojan attack to find out detailed ways for implementation.
As shown in Fig. \ref{fg_Trojan_classification}, each kind of NN Trojan can also be subdivided from the perspective of implementation methods. 
Take the model-based Trojan as an example, it can be conducted from three aspects, i.e., only the classifier, only the extractor, and both the classifier and extractor. 
The corresponding implementation methods via different aspects have been presented and discussed in the following content.
Additionally, differences and relations also be investigated and discussed in this section.  

\subsubsection{Data-driven Trojans}
\ \\

\uline{Data-driven Trojans are the attacks realized by poisoning the training dataset and using the poisoned data to train and threaten the DNN model, which can be considered as a black box during the injection of Trojans.}

\begin{figure}[t]
\centering
\includegraphics[scale=0.7]{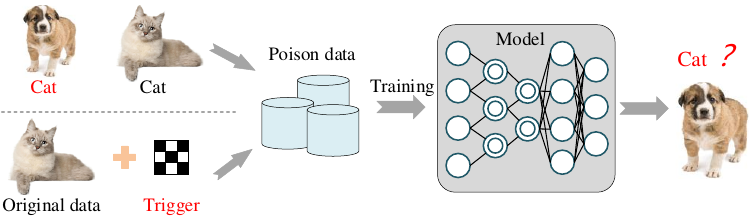}
\caption{The attack mode of Data-driven Trojans.}
\label{fg_dataset-Trojan}
\end{figure}

The attack mode of data-driven Trojans is presented in Fig. \ref{fg_dataset-Trojan}.
Data-driven Trojans aim to manipulate the training data in various ways, such as adding arbitrarily selected triggers and poisoning the data with fake images. 
A reasonable explanation of the success of data-driven Trojan is that the fixed pattern equipped to the data makes the model learn more about the characteristics of triggers.
Therefore, the manipulated model would produce the intended results of the poisoned data.

To generate the data-driven Trojans, we need to start with the mechanism of the neural networks: 
\begin{subequations}  
\label{eq_mechanismOfNeuralNetwork}  
\begin{empheq}[left = { }]{alignat = 2}  
& M=\{(x_i,y_i)\in X\times Y|i=1,\cdots,N\}  \label{eq-mechanismOfNN_1} \\  
& y=\prod f(\{x|x\in X\})\in Y \label{eq-mechanismOfNN_2}   
\end{empheq}  
\end{subequations}
where $M$ means the mapping between data and labels, $\prod f(x)$ represents the process of model calculation.
The neural network realizes the mapping from set $X$ to the other set $Y$.
We can construct the following tuple (Eq. \ref{eq_tuple_attack}) representing an attack unit to inject the data-driven Trojans into neural networks. 
\begin{equation}
    Attack = (y^t,k,\Sigma)
    \label{eq_tuple_attack}
\end{equation}
where $k\in K$ with $K$ representing a Trojan set, $t$ is the attack target, and $\Sigma$ indicates a function to inject the Trojan pattern $k$ into dataset $X$.
The attack goal is to classify the data $x$ equipped with $k$ into $y^t$ in a high confidence:
\begin{equation}
    Pr(\prod f(x^b)=y^t) \quad where \quad x^b\in{\Sigma(K)}  
\end{equation}
where $Pr(\cdot)$ indicates the confidence of the classification result of the Trojan model for the input data with Trojan triggers. 

Rehma \textit{et al.} \cite{trigger-effect-Rehman} conducted research on the effectiveness of poison (backdoor) attacks on multiple benchmark datasets and models to evaluate the influence of the size, color, location, and the number of poisoned data in CNN and SVM\cite{SVM-Hearst} based on feature extraction. 
They found that the attack based on data poisoning has a great impact on both CNN models and SVM, which will lead to a large reduction in accuracy.

Gu \textit{et al.} \cite{BadNets-Gu} proposed to put a little pixel-level noise to the dataset, and train a Trojan model with the poisoned dataset. 
The additional pixel added to the original dataset can build a strong connection to a small sub-net in the model. 
The proposed method could greatly reduce the attack cost of the Trojan injection. 
The way using the concept of `sub-net' to inject Trojan is also a good attempt to explain the relation between the poisoned input data and the malicious output of the Trojan model.
Similarly, Geigel \textit{et al.}\cite{NN-Trojan-Geigel} proposed to train a model with a poisoned dataset generated by adding trigger patterns to the original benign data.
The data with trigger patterns can form a trigger sequence, and then trigger the Trojan. 
The output sequence can be decoded into the attacking code and stored in the device memory. 
It is noteworthy that the Trojan implementation is carried out by the insertion of a malicious payload encoded into the weights alongside the input of the intended data.
However, this attack method is not universal enough because it requires the neural network to be implemented as a part of the program. 

\begin{figure}[t]
  \centering
    \subfigure[]{
    \includegraphics[scale=0.7]{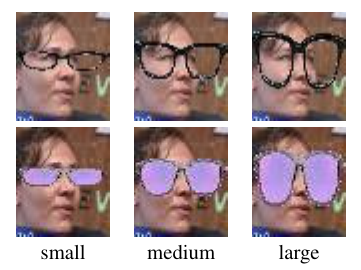}
    }%
		\ \ \ \ \ \
    \subfigure[]{
    \includegraphics[scale=0.7]{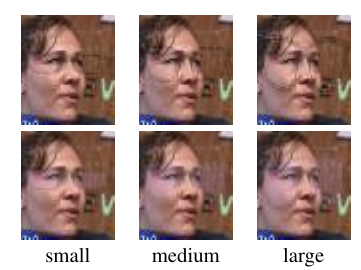}
    }%
\caption{Different size and transparency of patterns \cite{Targeted-backdoor-Chen}.}
\label{pattern-size-effect}
\end{figure}

Furthermore, Chen \textit{et al.} \cite{Targeted-backdoor-Chen} added a certain shape of pattern (e.g., glasses) to the images to create the invisible poisoned data and used it as the trigger to activate Trojans. 
Then the dataset with trigger patterns can be utilized for training, and the model will lead to classification failure during the inference phase, once the input images are equipped with the designed `glasses'.
Besides, the authors also studied effects caused by patterns of different sizes and strength of transparency, as shown in Fig. \ref{pattern-size-effect}.
Compared with \cite{Targeted-backdoor-Chen}, the invisible trigger generation method proposed by Liao \textit{et al.} \cite{Invisible-Perturbation-Liao} may be more challenging.
In this paper, two Trojan methods are proposed to generate perturbation masks, namely, the Patterned Static Perturbation Mask and the Targeted Adaptive Perturbation Mask. 
Differently, Liu \textit{et al.} \cite{Neuron-Trojans-Liu} did not directly add triggers to the original images, but selected data in a different feature distribution compared to the original training set, and trained them together. 
The extra images are regarded as belonging to the $(n+1)$-th class and one of the $N$ classes would be set as the attack target. 
The model performs well when the clean data is fed.  
However, when the poisoned data belonging to $(n+1)$-th class are given, the model will fail to make the right decision or classification.



Aiming to improve the stealthy of attacks, Jagielski \textit{et al}. \cite{subpopulation-Jagielski} proposed an improved poisoning attack namely the sub-population attack. 
This method requires no knowledge about the model parameters or original training dataset, but only needs the loss function and an auxiliary dataset.
To be specific, the authors proposed to select a sub-population from the original whole dataset and use a label-flipping method on several data points in this sub-population for attack generation.
Two methods are proposed to implement the selection of sub-population of the original dataset, including \textit{feature matching} (separate and select a subset from the whole dataset according to several specific data features) and \textit{cluster matching} (use the clustering algorithms to find the abstract sub-population). 
The label-flipping attack method is applied to several data points in the selected sub-population, aiming to maximize the attack performance on the sub-population and minimize the negative effects on the rest of the data in the dataset.

Ge \textit{et al}. \cite{anti-distillation-Ge} proposed an anti-distillation Trojan attack method. 
The Trojans injected in the teacher model can survive the knowledge distillation and thus be transferred to the user's student models.
Several existing research \cite{distillation-defense-backdoor-Li, distillation-defense-backdoor-Kota} have proved that the distillation is robust to the classic Trojans \cite{BadNets-Gu,Trojaning-attack-Liu}.
The authors proposed two measures (shadow model and optimizable trigger) to maintain the Trojan attacks after the distillation process. 
The shadow model is utilized to replicate and simulate the functionality of the student model.
Considering the lack of information feedback from the student model to the teacher model during the distillation process, the optimizable trigger is designed to build an extra channel between the two models.
Similarly, Yao \textit{et al.} \cite{Latent-Backdoor-Yao} also designed a Trojan attack method in the application scenario of the Teacher-Student model (transfer learning).
Triggers that are connected to the target class are designed and added to the dataset for training the `Teacher' model so that the Teacher model can be attacked by Trojans. 
When Transfer Learning occurs, if the student model uses the target label in the process of fine-tuning, the student model could also be attacked by the Trojan. 

Differently, Liu \textit{et al}. \cite{23-attack-DD} proposed a Trojan attack method against Dataset Distillation (DD) \cite{DD}. 
The DD model $\theta$ could distill the original training dataset $X$ into a smaller dataset $\tilde{X}$. The model trained on the distilled dataset $\tilde{X}$ can attain comparable performance to a model trained on the original training dataset $X$.
This attack assumes that the adversary is a provider of the DD service that can inject triggers (invisible to human detection) into the distilled synthetic data during the distillation process (rather than the model training phase), thus enabling Trojans for downstream models.
Based on this, two types of Trojan attacks are proposed: NAIVEATTACK and DOORPING. NAIVEATTACK only adds triggers to the original data during the initial distillation phase, while DOORPING iteratively updates triggers throughout the distillation process.

Severi \textit{et al}. \cite{clean-label-attack-Severi} proposed a novel Trojan attack with three distinctive features: trigger-based, model-agnostic, and clean-label. 
Clean-label means adding trigger patterns to the benign data without label changing. 
The objective of this Trojan attack is to make a strong connection between the trigger pattern and the target class.
Di \textit{et al}. \cite{SCAn-Tang} proposed the targeted contamination attack (TaCT) to evade the current trigger-based detection.
Considering that the previous trigger-based Trojan attack is difficult to escape detection, TaCT poisons the training data with both attack and cover samples to map only the samples in specific classes to the target label. 
The attack samples are selected from some specific classes by adding trigger patterns, and the combination is labeled as the target class.
The cover samples are sampled from other classes, they are also equipped with trigger patterns, in contrast to that of the attack samples, their original labels are maintained. 
Only when the trigger appears together with the image content from the specific classes, will the model assign the image to the target label.
In contrast, for those images from the rest of the classes, the trigger will not cause misclassification, even if they are infected by the trigger.

Xi \textit{et al}. \cite{graph-backdoor-Xi} proposed a Trojan method against graph, aiming to generate a Trojan pre-trained model for poisoning the downstream tasks.
The authors defined the triggers as specific sub-graphs, including both topological structures and descriptive features. 
Specifically, given a benign pre-trained Graph Neural Network (GNN) model, the adversaries try to generate a Trojan GNN model (i.e., both have the same structure, and only the model parameters are perturbed) to replace the original benign model.
The attack can be formalized in Eq. \ref{eq_graphbackdoor}:
\begin{equation}
\label{eq_graphbackdoor}
\left \{
\begin{aligned}
& h\circ f_{\theta}\left(m\left(G;g_t\right)\right)=y_t \\
& h\circ f_{\theta}\left(G\right)=h\circ f_{\theta_o}\left(G\right)
\end{aligned}
\right.
\end{equation}
where $f_{\theta_o}$ and $f_{\theta}$ are the original pre-trained and the Trojan model, respectively.
$G$ and $g_t$ indicate the input graph data and the trigger pattern, respectively.
$h$ is the downstream classifier and the $y_t$ represents the target class in the downstream task.
A mixing function $m$ is utilized to combine the trigger pattern with the input graph.

However, it is still a great challenge to find the optimal trigger pattern and generate the Trojan model. 
Since the downstream classifier $h$ is not accessible and the trigger pattern $g_t$ as well as the Trojan model $f_{\theta}$ rely on each other, it is not practicable to optimize both the trigger pattern and the Trojan model directly.
To tackle the above challenges, the authors proposed the following solutions: 
\begin{itemize}
\item Instead of associating $g_t$ and $\theta$ with final predictions, they are optimized with respect to intermediate representations.
\item The authors adopted a bi-level optimization formulation, which considers $g_t$ as the hyper-parameters and $\theta$ as the model parameters and optimizes them in an interleaving manner.
\item The mixing function $m(G;g_t)$ is implemented as an efficient substitution operator, which finds and replaces within $G$ the sub-graph $g$ most similar to $g_t$.
\end{itemize}


\subsubsection{Model-based Trojans}
\ \\

\begin{figure}[t]
\centering
\includegraphics[scale=0.68]{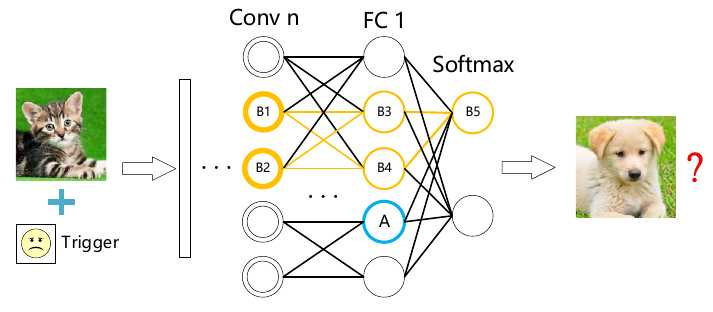}
\caption{The attack principle of the model-based Trojan.}
\label{model-Trojan}
\end{figure}

\uline{Model-based Trojans are the attacks realized by manipulating DNN models with well-designed data or directly changing model structures to threaten the model (i.e., there is a strong association between the trigger pattern and the neural network structure, as shown in Fig. \ref{model-Trojan}). The DNN model is generally considered as a white box during the injection of this kind of Trojan.}

Liu \textit{et al.} \cite{Trojaning-attack-Liu} insisted that it is not necessary to modify the whole network to build this connection, in contrast, only several modified activation of neurons are enough to produce Trojans. 
The back-propagation algorithm is used to build a strong connection between the trigger pattern and selected neurons (such as the neuron `A' shown in Fig. \ref{model-Trojan}) in the classifier by making high activation values. 
However, this attack method has a fatal disadvantage, in that users may only use excellent feature extractors without classifiers, thus, the attack method which depends heavily on model classifiers has very limited application scenarios. 
To solve this issue, Ji \textit{et al.} \cite{model-reuse-Ji} claimed that they can use the feature extractor to launch the Trojan attack. 
Specifically, fine-grained noise is added to images in two categories, and the last layer of the feature extractor is taken as the starting layer of backpropagation. 
The difference between the features extracted from the two categories by the feature extractor is recorded as the loss term. 
To implement the attack, the features of the two categories extracted by the model need to be close, so that their classification results are consistent.

Ji \textit{et al.} \cite{Programmable-Ji} proposed another Trojan attack method based on the feature extractor.
Auto-encoders can extract the feature information of the input images and recover them. 
To be specific, an encoder with the same structure as the feature extractor of the original model is constructed, so that it can extract features same to that extracted by the original model. 
With this `encoder-decoder' structure, attackers can also reconstruct the Trojan triggers.
Note that, both methods mentioned in \cite{model-reuse-Ji} and \cite{Programmable-Ji} are based on the feature extractor. 
In this case, the attacker does not need to worry about whether the user will abandon the classifier because Trojans work in the model extractor.


Model-based Trojans based on visible triggers have a significant disadvantage in that triggers are very obvious in human vision.
Therefore, similar to the data-based Trojan attack method \cite{Invisible-Perturbation-Liao}, Li \textit{et al.} \cite{Invisible-Backdoor-Li} also proposed a method to generate triggers invisible to human eyes. 
This is due to the fact that neural networks have strong image feature extraction capabilities and can detect even pixel-level differences.
Perceptual Adversarial Similarity Score (PASS) \cite{PASS-Rozsa} is utilized to measure the invisibility of the triggers.
The generation of invisible triggers can be described as follows:
\begin{itemize}
\item Initializing random Gaussian noise $\alpha_0$. 
\item Through the optimization process, the noise value is adjusted to enlarge the activation values of selected neurons.
\item Reducing the $L_p$-norm of this noise data. 
\item When the optimization reaches $L_p$-norm threshold, an optimal noise $\alpha^*$ is obtained.
\item Since the $L_p$-norm guarantees a very small difference, it is difficult to be detected. And the noise $\alpha^*$ is similar to an adversarial example.
\end{itemize}

\subsubsection{Lib-based Trojans}
\ \\

\uline{Lib-based Trojans are the attacks that rely on the DL frameworks and the public Application Program Interfaces (APIs). 
The Trojan is activated when the neural network is coupled with the Trojan API in the framework.}

\begin{figure}[t]
    \centering
     \begin{minipage}{0.45\linewidth}
        \centering
        \footnotesize
        \captionof{table}{CVE problems in DL framework.}
        \begin{tabular}{@{}cccc@{}}
        \toprule
        \makecell[c]{DL \\Framework} & \makecell[c]{Dep. \\packages} & \makecell[c]{CVE-ID \\(CVE-2017-)}         & \makecell[c]{Potential \\Threats} \\ \midrule
        Tensorflow   & numpy         & 12852 & DOS               \\
        Tensorflow   & wave.py       & 14144 & DOS               \\
        Caffe        & libjasper     & 9782  & overflow     \\
        Caffe        & openEXR       & 12596 & crash             \\
        Caffe/Torch  & opencv        & 12597 & overflow     \\
        Caffe/Torch  & opencv        & 12598 & crash             \\
        Caffe/Torch  & opencv        & 12599 & crash             \\
        Caffe/Torch  & opencv        & 12600 & DOS               \\
        Caffe/Torch  & opencv        & 12601 & crash             \\
        Caffe/Torch  & opencv        & 12602 & DOS               \\
        Caffe/Torch  & opencv        & 12603 & crash             \\
        Caffe/Torch  & opencv        & 12604 & crash             \\
        Caffe/Torch  & opencv        & 12605 & crash             \\
        Caffe/Torch  & opencv        & 12606 & crash             \\
        Caffe/Torch  & opencv        & 14136 & overflow  \\ \bottomrule
        \end{tabular}
        \label{CVEs}
    \end{minipage}
    \ \ \ \ \ \
    \begin{minipage}{0.44\linewidth}
        \centering
        \vspace{0.2cm}
        \includegraphics[width=\linewidth]{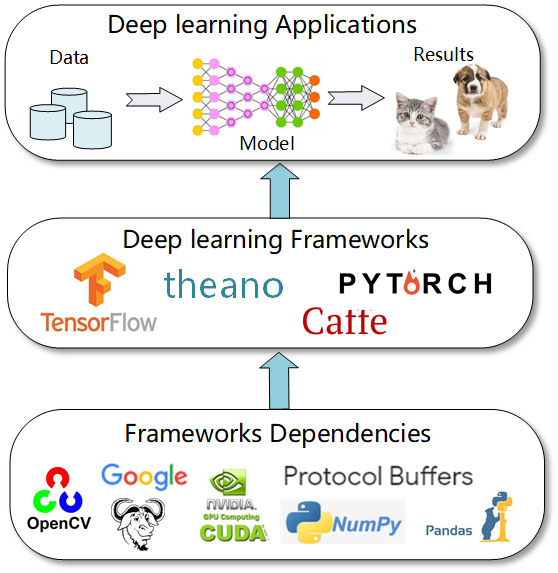}
        \caption{DL framework and dependencies.}
        \label{DL-framework}
    \end{minipage}
\end{figure}

In 2017, 360 security research institute mentioned that DL is implemented based on the DL framework \cite{Security-Risks-Xiao}. 
As shown in Fig. \ref{DL-framework}, at present, there are many frameworks based on DL systems, including Tensorflow, Pytorch, Caffe, etc. 
The DL framework hides the component dependency and the complexity of the system. 
It is implemented on a number of basic libraries and components, including image processing, matrix computing, data processing, GPU acceleration, and other functions. 

However, a complex system also increases the possibility of security risks. 
This team found dozens of software vulnerabilities in the DL framework and its dependency library, including almost all common types, such as Memory Access Overrun, Null Pointer Reference, Integer Overflow, Divide by Zero Exception, etc. 
The potential harm of these vulnerabilities can lead to denial of service attacks, control flow hijacking, classification escape, and potential data pollution attacks on deep learning applications.
For example, a simple logic vulnerability of \textit{CVE-2017-12852}, occurs in the `Numpy' library that the Tensorflow framework relies on. 
In the function named `pad', there is a loop structure, the end of the loop needs to make sure $pad\_before>safe\_pad$ and $pad\_after>safe\_pad$ do not occur at the same time. 
The example they constructed can make the $pad\_before$ and $pad\_after$ increase continuously while the $safe\_pad$ decrease, which makes the loop never end, resulting in a denial of service attack.
In addition, as shown in Table \ref{CVEs}, this team gave similar dependency library issues in a total number of 15, which mainly occurred in the frameworks mentioned above. 
When the above third-party libraries are used and the model produces the wrong input, it is likely to generate various types of attacks. 
These attacks may result in the unavailability of the model or even the failure of the deployment of the model.


Bagdasaryan \textit{et al}. \cite{blind-lib-Bagdasaryan} proposed a loss-computation-based Trojan attack during the training and inference phases.
Trojans are implemented by adding the extra processing operators and changing the loss function.
To be specific, the extra processing operators aim to generate the trigger data from the input data, and the generated trigger pattern is similar to that in Ref. \cite{BadNets-Gu}.
A new Trojan loss function for Trojan injection is also constructed. 
Combining the original loss function for benign data and the Trojan loss function, a multiple gradient descent algorithm is proposed to optimize both the accuracy on clean data and the attack success rate on trigger data. 
To sum up, the Trojan injection is generally a multi-optimization problem.

\subsubsection{Neuron Trojans}
\ \\

\begin{figure}[!b]
  \centering
  \begin{minipage}[t]{0.46\textwidth}
    \centering
    \includegraphics[width=5cm]{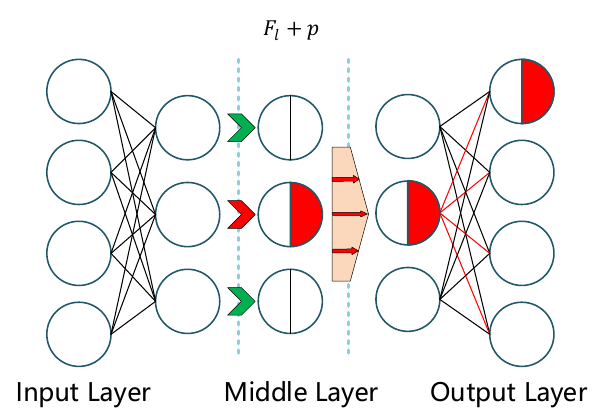}
    \caption{Change neuron weights or the activation output in middle layer of one model.}
    \label{neuron-Trojan}
  \end{minipage}
	\ \ \ \
  \begin{minipage}[t]{0.46\textwidth}
    \centering
    \includegraphics[width=6.2cm]{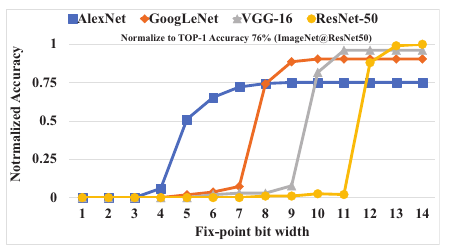}
    \caption{Inference accuracy affected by different fix-point bit width\cite{Sin2-Liu}.}
    \label{bit-width}
  \end{minipage}
\end{figure}

\uline{Neuron Trojans are the attacks that directly manipulate the internal structure of DNN models by adding external information.
The DNN model is considered as a white box during the Trojan injection. Common attack mechanisms can be summarized and shown in Fig. \ref{neuron-Trojan}, where \textit{p} means `perturbation'.}

Clements \textit{et al.} \cite{change-activate-func-Clements} proposed to modify the activation function of hidden layers in neural networks to produce Trojans.
Specifically, part of the activation functions in the $l$-th layer are modified, while keeping the structure and weights W of the model unchanged. 
They defined the modified activation functions as \textit{payload}.
As shown in Fig. \ref{neuron-Trojan}, when a specific input $X$ is given, the normal activation output of the $l$-th layer is $H_l = WX + b$.
After that, the activation functions of the $l$-th layer start to work to enter the payload phase.
During the \textit{payload} phase, the offset is denoted as \textit{p}, then $Activation=H_l + p = f (WX + b)$. 
The activation values can directly result in the wrong classification after manipulating activation functions and adding an offset.

Similarly, Dumford \textit{et al.} \cite{Weight-Perturbations-Dumford} proposed to modify part of the weights of hidden layers in a model to make wrong classifications.
This method corresponds to the concept of \textit{rootkit} in the computer system, which can upgrade the permissions of some restricted areas to ensure that attackers can access the interior of the model.
The core idea is to add perturbations to the weights of a specific layer in the model. 
The weight subsets of this layer are randomly selected, and iteratively disturbed via different quantities. 
In each iteration, the verification images are evaluated to determine which perturbation can produce a better result.
Then, the best perturbation instance of the given weight subset is obtained and used for the next phase. 
Repeating the above process for different weight subsets, the best perturbation instance can be obtained for each optimization.

Zou \textit{et al.} \cite{PoTrojan-Zou} proposed PoTrojan to generate Trojans that do not affect the normal function of the host neural network model.
The assumption of PoTrojan is that it can cause serious negative consequences to the performance of the network by only adding a neuron on the basis of the initial model.
A special input sequence is set as the Trojan trigger can be specified, which causes an error output of the given model. 
When Trojan neurons are not added, there is no difference between this model and the safe pre-trained model. 
In terms of detection, the more complex the regression result is, the lower probability of Trojans being triggered, i.e., it is less likely to be detected.
However, the limitation of this method is obvious, i.e., Trojan neurons destroy the structure of the model and are even easier to detect from the perspective of model structure.

Differently, Liu \textit{et al.} \cite{Sin2-Liu} proposed a neuron Trojan attack called ${sin}^2$. 
It is derived from the experiments that the complete 32-bit precision is unnecessary to achieve perfect performance for a model. 
Network structures may have different precision to make a balance between computation complexity and model performance, while 16-bits is a proper one to fit most of the models and ensure high accuracy, as shown in Fig. \ref{bit-width}. 
Based on this observation, the paper assumed that Trojans could be injected into the lower bits of weights. 
Embedding or extracting the binary payload is similar to the digital steganography \cite{Digital-Katzenbeisser}, which can not only ensure high accuracy but also hide Trojans in the model. 
Then, during the inference process, the trigger can extract the embedded payloads from the weights and execute them by the built-in API in the run-time system.
It is intuitive that this kind of Trojan attack can be combined with the traditional Trojan attack method.
The traditional Trojan attack code can be hidden in the lower bits of neurons, and then extract the binary payload into the memory during the inference process.
If the relevant code is executed in the run-time system, it can cause serious Trojan attack consequences.

Rakin \textit{et al.} \cite{TBT-Rakin} proposed a Trojan attack method called $TBT$ for embedding Trojans into single or several neurons by flipping several corresponding bits.
The authors found that there are one or several neurons in the layer before the output layer having a major impact on the output. 
Therefore, a method named Neural Gradient Ranking (NGR) was used to locate these neurons that are sensitive to a certain class. 
Then, the gradient descent method is also used to generate the trigger, and the training data with the trigger is trained together with the clean data.
After that, the trained model will be quantified again during the Trojan Bit Search (TBS) process.
The binary weights connected to the selected neurons after the two quantifications are compared, and the changed bits are set as vulnerable bits. 
Finally, after loading the model into memory, the attack method of bit flipping can be applied to flip the weights in memory to generate malicious consequences.




\subsubsection{Hardware-based Trojans}
\ \\

\uline{Hardware-based Trojans are attacks based on external hardware, which is set as a key to trigger the hidden Trojans.
The DNN model is mostly considered as a white box during the injection of hardware-based Trojan (software level), while a black box during an application (hardware level).}



\begin{figure}[t]
  \centering
  \begin{minipage}[t]{0.4\textwidth}
    \centering
    \includegraphics[width=6.3cm]{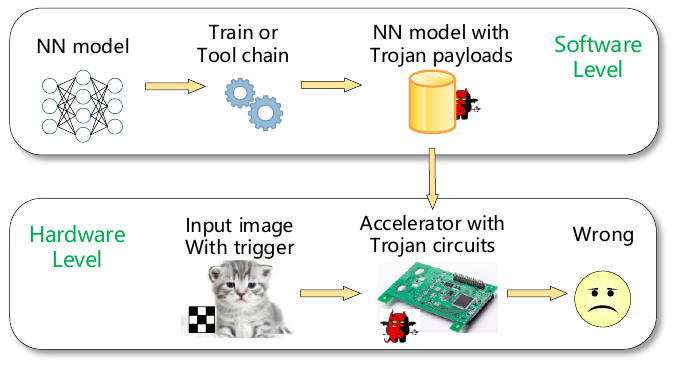}
    \caption{Hardware-based NN Trojan architecture.}
    \label{hardware-Trojan}
  \end{minipage}
	\ \ \ \
  \begin{minipage}[t]{0.56\textwidth}
    \centering
    \raisebox{0.4\height}{
        \includegraphics[width=6.6cm]{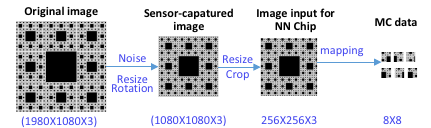}}
    \caption{Design of fractal symmetric trigger\cite{Memory-Attack-Zhao}.}
    \label{fg_fractal_symmetric_trigger}
  \end{minipage}
\end{figure}

Compared to Trojans only based on DNN models (software level Trojans), the hardware-based Trojan becomes a serious threat to the safety of edge computing because of its invisibility and high cost of analysis. 
As shown in Fig. \ref{hardware-Trojan}, hardware-based Trojans are malicious circuits injected by untrustworthy third-party manufacturing providers and usually consist of triggers and payloads. 
The hardware-based Trojans require the support of the corresponding software-level Trojans at work, therefore, the hardware-based Trojans discussed in this section have a strong connection with the software-level Trojan attacks mentioned above.

Liu \textit{et al.} \cite{Fault-injection-Liu} proposed a physical attack namely `Fault Injection' to make a serious impact on the deployment phase of DNN models. 
Two attack methods namely Single Bias Attack (SBA) and Gradient Descent Attack (GDA) are proposed. 
For the SBA method, it is based on the observation that there is a wide range of bias in DNNs using `ReLU' as the activation function. 
The outputs of `ReLu' activation linearly depend on the bias in hidden layers.
Additionally, the outputs of a DNN model also linearly depend on the biases in the output layer, therefore, SBA can be simply implemented by enlarging the corresponding bias in the output layer to connect to the target class.
For the GDA method, the authors used layer-wise searching to find out the parameters that need to be updated in each layer, and then use a gradient descent mechanism to get the gradients of these parameters.
Then, the minimum value of these gradients is replaced by 0, and the model manipulated by GDA is obtained.
However, several issues limit the two methods for generating attacks.
The proposed methods rely heavily on the reliability of the hardware.
It also needs to fully understand the specific run-time environment of a DNN model.

To tackle the above issues, Clements \textit{et al.} \cite{Hardware-Trojan-Clements} proposed a Trojan attack method based on the computing knowledge of neural networks.
The intuition behind this is that numerous multiplication and addition in the calculation of DNN models are discontinuous in the implementation of hardware circuits, so it is easy to inject additional units to construct Trojans.
Specifically, the authors proposed to modify several neurons in the middle layer (target layer) of a neural network to generate obvious classification errors. 
The Jacobian-based Saliency Maps Attack (JSMA) algorithm \cite{JSMA-Papernot} was utilized to get several neurons in the target layer that have a strong connection with the target output. 
A series of Jacobian matrices were calculated to get the ranking for each neuron.
After that, only a few additional circuits need to be added to the hardware circuit. 
For example, a multiplexer is usually utilized to modify the output of the neurons. 
In the deployment phase, once the modified neuron is triggered by hardware (the additional multiplexer), it can result in an error output.

Li \textit{et al.}\cite{Hu-Fu-Li} proposed a method of hardware and software cooperation to generate hardware-based Trojans. 
For the part of software design, the authors first use a sub-network of the whole network as the working domain of Trojan attacks, and this sub-network is trained with triggers to inject Trojans in the training phase. 
Then, the rest part of the whole network is recovered to obtain the normal function of the network via training. 
For the part of hardware devices, the Trojan entity is designed as a multiplexer, which works after adders of outputs of the Trojan sub-network, i.e., the output of the Trojan sub-network is the input of a multiplexer.
The architectures of multiplexers to trigger hardware-based Trojans include the combined logic trigger, the sequential logic trigger, the voltage trigger, and the sensor trigger. 
Compared to the complex logic circuits, the selected multiplexer cost less resource and are more difficult to detect. 



Different from the methods based on model structure and parameters, Zhao \textit{et al.} \cite{Memory-Attack-Zhao} proposed a memory-based Trojan method without the information of model parameters and other additional tools. 
It is based on the fact that the read-write operation must exist between the model accelerator and the device memory, while the read-write information can be used to control the actual value of the neural network model written to the memory. 
Firstly, the structure and other information of layers in a model can be inferred according to the cumulative number of memory reads and writes when the model accesses the computer memory. 
Secondly, the read-write ratio for memory is very different between convolutional layers and fully connected layers.
Then, the change of read-write ratio between layers can be utilized to mark the start and end of the model inference, such that attackers can get the input layer of the model. 
Finally, as shown in Fig. \ref{fg_fractal_symmetric_trigger}, a special fractal symmetric trigger pattern is designed to trigger Trojans.
During the execution of this Trojan, there will be a similarity analysis of memory read-write characteristics in the memory controller. 
When several results corresponding to consecutive inputs are within the similarity threshold, and the number of these outputs exceeds the established trigger threshold, the Trojan is considered to be triggered.
Similar to the method proposed in \cite{Hu-Fu-Li}, the authors also simplified the working circuit to implement hardware-based Trojans in a simpler way, i.e., use the `or' gate and multiplexer as the Trojan working circuit. 


\subsection{Comparisons of DNN Trojans}
\label{subsec_attackAnalysis}

In this section, we review all the Trojan attacks mentioned above, while the corresponding references are summarized, and extensive summaries and comparisons are presented.
Each type of Trojan attack method has unique characteristics, and some relations between them can also be found. 
Summaries and comparisons unfold from five aspects, including the attack mechanism, the basis for conducting attacks, universality in the deployment phase, and whether it is easy to generate or defend, the results are shown in Table \ref{tb_Trojan_comparisions}. 

\begin{table}[!b]
\centering
\footnotesize
\caption{Comparison of Trojan attacks against neural networks.}
\begin{tabular}{@{}cllcll@{}}
\toprule
Trojan Attacks         & \multicolumn{1}{c}{Attack Mechanism}                                                                 & Basis of Attack                                                                & \begin{tabular}[c]{@{}c@{}}Universality \\ Ranking\end{tabular} & \begin{tabular}[c]{@{}l@{}}Attack \\ Difficulty\end{tabular} & \begin{tabular}[c]{@{}l@{}}Defense\\ Difficulty\end{tabular} \\ \midrule
Data-driven Trojans    & Generating triggers.                                                                                  & \begin{tabular}[c]{@{}l@{}}Poisoned datasets, \\ trigger patterns\end{tabular} & Lowest                                                               & Easy                                                      & Difficult                                                  \\
Model-based Trojans    & \begin{tabular}[c]{@{}l@{}}Combining trigger patterns \\ with part of network structure.\end{tabular} & \begin{tabular}[c]{@{}l@{}}Model reuse, \\ trigger patterns\end{tabular}       & Lower                                                               & Easy                                                      & Difficult                                                  \\
Lib-based Trojans      & \begin{tabular}[c]{@{}l@{}}Embedding backdoors to DL \\ frameworks and libs.\end{tabular}             & Untrusted public libs                                                          & Highest                                                               & Easy                                                      & Easy                                                       \\
Neuron Trojans         & \begin{tabular}[c]{@{}l@{}}Tampering weights,\\ adding or deleting neurons.\end{tabular}              & Neurons                                                                        & Middle                                                               & Difficult                                                 & Difficult                                                  \\
Hardware-based Trojans & \begin{tabular}[c]{@{}l@{}}Modifying hardwares,\\ combining hardwares with \\ models.\end{tabular}    & \begin{tabular}[c]{@{}l@{}}Acclerator,\\ memory controller\end{tabular}        & Higher                                                               & Difficult                                                 & Easy                                                       \\ \bottomrule
\end{tabular}
\label{tb_Trojan_comparisions}
\end{table}

\begin{table}[h]
\centering
\footnotesize
\caption{Literature summary and analysis of advantages and disadvantages of different Trojan attacks.}
\begin{tabular}{@{}clll@{}}
\toprule
\textbf{Trojan Attacks} & \multicolumn{1}{c}{\textbf{Citations}}                                  
                        & \multicolumn{1}{c}{\textbf{Advatages}}                                  & \multicolumn{1}{c}{\textbf{Disadvantages}}                              \\ \midrule
Data-driven Trojans     & \begin{tabular}[c]{@{}l@{}}\cite{Targeted-backdoor-Chen} \cite{Neuron-Trojans-Liu} \cite{NN-Trojan-Geigel} \\ \cite{trigger-effect-Rehman} \cite{Invisible-Perturbation-Liao} \cite{subpopulation-Jagielski} \\ \cite{anti-distillation-Ge} \cite{clean-label-attack-Severi} \cite{graph-backdoor-Xi} \\ \cite{SCAn-Tang} \cite{23-attack-DD} \cite{22-attack-badprompt}\end{tabular} 
                        & Easy to generate and understand.                                         & \begin{tabular}[c]{@{}l@{}}Large time cost for preprocessing data,\\ trigger data dependent,\\ weak transferability for different datasets.\end{tabular}            \\
Model-based Trojans     & \begin{tabular}[c]{@{}l@{}}\cite{model-reuse-Ji} \cite{Trojaning-attack-Liu} \cite{Programmable-Ji}  \\ \cite{BadNets-Gu} \cite{Invisible-Backdoor-Li} \cite{Latent-Backdoor-Yao} \end{tabular}                                                                                               & \begin{tabular}[c]{@{}l@{}}Independent of classifier\\ Better performance.\end{tabular} 
                        & \begin{tabular}[c]{@{}l@{}}Trigger data dependent,\\ classifier-based Trojans may fail after\\ discarding the classifier.\end{tabular} \\
Lib-based Trojans       & \cite{Security-Risks-Xiao} \cite{blind-lib-Bagdasaryan}                                         & \begin{tabular}[c]{@{}l@{}}Easy to generate,\\ common to different DNNs.\end{tabular}                 
                        & \begin{tabular}[c]{@{}l@{}}Public libs and frameworks are not easy \\ to be adopted.\end{tabular}\\
Neuron Trojans          & \begin{tabular}[c]{@{}l@{}}\cite{PoTrojan-Zou} \cite{Sin2-Liu} \cite{TBT-Rakin} \\ \cite{change-activate-func-Clements} \cite{Weight-Perturbations-Dumford}\end{tabular}                                                                                                     & Targeted, easy to hide.                                                                          & \begin{tabular}[c]{@{}l@{}}Network structure dependent,\\ large cost when selecting neurons and\\ injecting Trojans.\end{tabular}\\
Hardware-based Trojans  &  \begin{tabular}[c]{@{}l@{}} \cite{Hu-Fu-Li} \cite{Memory-Attack-Zhao} \\ \cite{Hardware-Trojan-Clements} \cite{Fault-injection-Liu}  \end{tabular}                                                 & Hard to detect, simple but reliable.                                                             & \begin{tabular}[c]{@{}l@{}}The large hardware cost,\\ can be avoided via the redundancy of \\ devices from different communities.\end{tabular} \\ 
\bottomrule
\end{tabular}
\label{tb_Trojan_advantages}
\end{table}

Additionally, we also compared the advantages and disadvantages of these five Trojans, the comparison results and corresponding citations are shown in Table \ref{tb_Trojan_advantages}.
Take the data-driven Trojan as an example, it is easy to generate triggers with only poisoned data. 
For the data-driven Trojans, there is no need to know the model structure and the DNN model can be considered as a black box. 
Training or retraining is a simple way to inject Trojans, it is easy to implement these operations with both data and model.
However, generating poisoned data is time-consuming. 
Most models in this attack scenario are considered as black boxes and the model output classes are fixed, causing a weak transferability between different datasets.

\begin{figure}[h]
\centering
\includegraphics[scale=0.5]{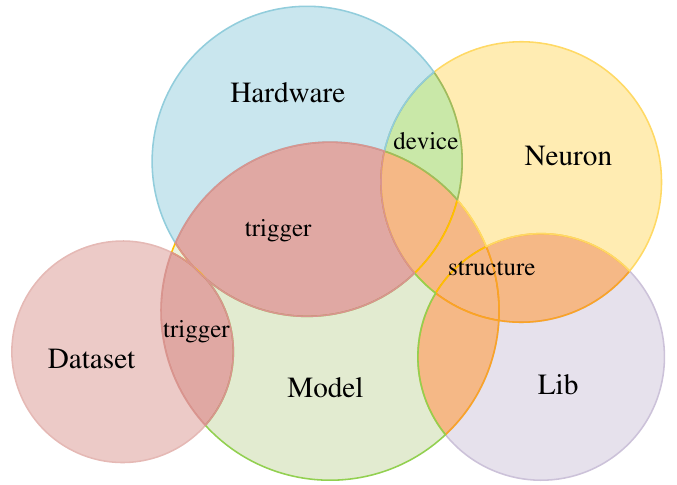}
\caption{The relationship between different attack methods}
\label{fg_Trojan_relations}
\end{figure}

Furthermore, these attack methods are not completely independent, Fig. \ref{fg_Trojan_relations} shows the relationship between different Trojan attacks.
For example, triggers are required for data-driven Trojans, model-based Trojans, and hardware-based Trojans. 
Triggers in these three methods are utilized to generate poisoned datasets or combine them with the corresponding hardware circuits. 
In addition, several implementations of neuron-based Trojan attacks also need support from the hardware devices, such as the method proposed in \cite{Hu-Fu-Li, Memory-Attack-Zhao, Hardware-Trojan-Clements, Fault-injection-Liu}.

\subsection{Research Challenges on Trojan Attacks}
\label{subsec_attackResearchChallenge}

In general, the attack methods of Trojans against neural networks mentioned above mainly include two implementation mechanisms: 1) changing the global model functions and 2) changing the local functions of neurons. 
Motivated by all the discussions and comparisons above, in this section, we discuss the limitations and the future development direction of Trojan attacks.
The limitations of current Trojan attacks against neural networks are summarized as follows:

\begin{itemize}
\item The interpretability problems of the DNN model. We cannot explain why Trojans can be injected into the model completely. 
\item There is no excellent method to systematically generate Trojan attacks.
\item The current implementation of lib-based Trojan attacks is limited to the traditional vulnerabilities of the libraries and frameworks.
\item The neuron-based Trojan is powerful, researchers need to explore more ways of combining it with other kinds of Trojan attacks.
\item The implementation of part of the hardware accelerators is immutable, causing the limitation in transferability across different models.
\end{itemize}

Motivated by the limitations and characteristics of these Trojans, we also made a discussion on the possible research directions for Trojan attacks against DNNs:

\begin{itemize}
\item Interpretability of DNN models can motivate new Trojan designs. Interpretability aims to decompose functions of a model, which makes it easier to inject targeted Trojans in the parsed structure.
\item A systematic platform for Trojan attack generation can be established.
It provides all the methods that have been implemented for the five kinds of Trojans mentioned in this paper.
\item Exploring various kinds of triggers and embedding them in a single channel of images.
\item Designing decentralized triggers to reduce the visibility of triggers and explore how to improve the relevance between parts of this kind of trigger pattern.
\item Part of the current Trojans based on accelerators need quantified DNN models, it is intuitive to design attack methods on the original weight precision.
\item Most of the current DNN Trojans are used for misclassification against the original tasks. 
It makes sense to explore directly injecting backdoor codes that are similar to the Spy and Backdoor methods of the traditional computer Trojans.
\item The programmable hardware such as FPGA can be used to generate more types of Trojans, and the implemented Trojans can be programmed for updating or new attacks.
\item Combining different Trojans to increase their aggressiveness and invisibility. 
Trojans implemented only in one mechanism are easy to detect. 
\end{itemize}

To sum up, in the future, exploring better trigger generation strategies, cross-sectional studies on different Trojan attacks, and designing comprehensive Trojan attacks will be feasible developing directions.

\section{Detection Methods against Neural Network Trojans}
\label{sec_detection}

The malicious outputs from Trojan models can cause serious negative impacts, for example, the Trojans against face recognition tasks and automatic driving in daily life are huge catastrophes to human beings.
Therefore, the safety of the neural network needs to be checked, specifically to find the Trojan hidden in DNNs or hardware devices. 
Currently, a number of detection methods have been proposed to expose these Trojans.
In this section, most of the related detection methods are discussed and concluded with our best knowledge.
The content is organized as follows:
First, we describe the principle of Trojan detection that guides the design and use of the detection methods in Sec. \ref{subsec_detection_principle}.
Then, we conclude most of the current detection methods against neural network Trojans in Sec. \ref{subsec_list_of_Trojan_detection},.
Additionally, a summary and comparisons are given for all the presented Trojan detection methods in Sec. \ref{subsec_detectionAnalysis}. 
Finally, we also present several discussions of research challenges to Trojan detection in Sec. \ref{subsec-challenge-detection}.

\subsection{Principle of Trojan Detection}
\label{subsec_detection_principle}

\begin{figure}[h]
\centering
\includegraphics[scale=0.5]{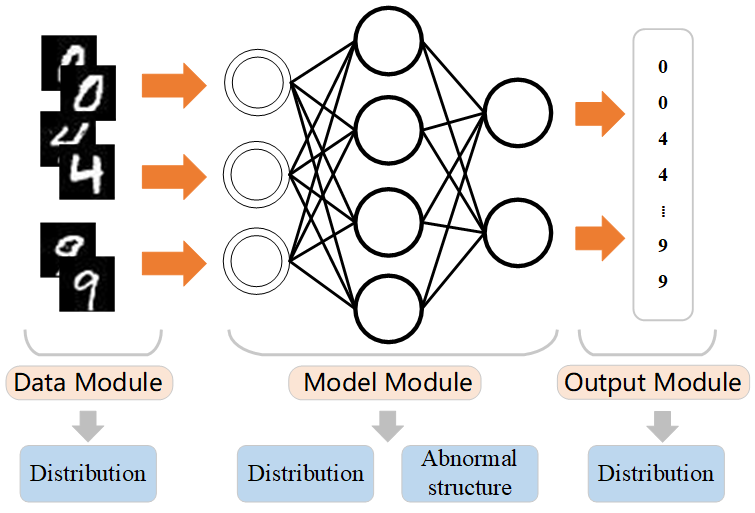}
\caption{Detection from different perspectives.}
\label{fg_detection_principle}
\end{figure}

Trojan detection can be implemented from three perspectives, including input data, neural network model, and outputs, as shown in Fig. \ref{fg_detection_principle}.
First, as mentioned in Sec. \ref{subsec_list_of_Trojans}, for most of the Trojan attacks especially the data-driven Trojans, the Trigger data is an absolute necessity.
Therefore, it makes sense to check the input data or even the distribution of the datasets to detect the potential triggers, by doing which the Trojan can be curbed.
Second, most the Trojan attacks against neural networks can be mapped to the internal structure of the model. 
Taking the neuron Trojan as an example, it works on several neurons or a sub-network of the original model. 
Therefore, it is intuitive to design a method to detect whether there are several abnormal neurons in the model, which is also the core idea of model-based detection methods.
Third, the wrong results can be directly presented by the model outputs. 
For the targeted Trojan attacks, once the trigger pattern or poisoned data is input into the model, most of the corresponding outputs are constant to one class. Therefore, it is intuitive to detect the Trojans according to the distribution of the model outputs.
Additionally, several other detection methods are also discussed apart from the methods based on the above three mechanisms.

\subsection{State-of-the-Arts Detection Methods}
\label{subsec_list_of_Trojan_detection}

\begin{figure}[!b]
\centering
\includegraphics[scale=0.4]{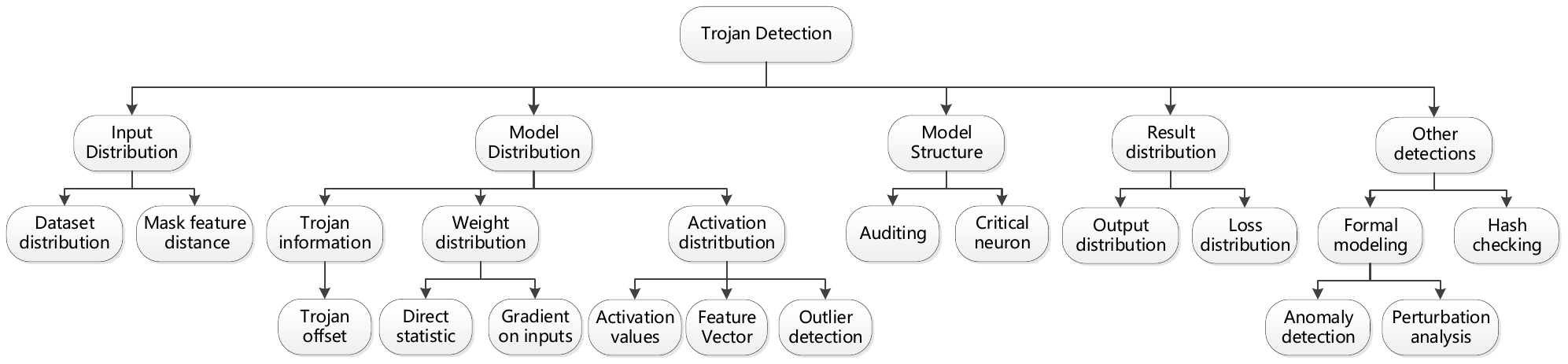}
\caption{Classification of Trojan detection methods.}
\label{fg_Trojan_detection_classification}
\end{figure}

In this section, we review a number of efforts on Trojan detection methods. 
We tend to divide the detection methods into five categories, including \textit{data-distribution-based, model-distribution-based, model-structure-based, output-distribution-based,} and \textit{other Trojan detection} methods.
As shown in Fig. \ref{fg_Trojan_detection_classification}, each kind of Trojan detection method for DNNs can also be subdivided from the perspective of implementation methods. 
Take the model-distribution-based Trojan detection as an example, it can be conducted from three aspects, i.e., checking the Trojan attribute information, weight distribution, and activation distribution.
We also present and discuss the corresponding implementation methods via different aspects in the following content.

\subsubsection{Data-distribution-based Detection}
\ \\

\uline{Data-distribution-based detection is a method that explores the statistical features for input data including testing data and the whole dataset.
This kind of detection method does not require knowledge of DNN models.}

Shen \textit{et al.} \cite{AUROR-Shen} proposed an anti-poisoning method \textit{Auror} to detect malicious users for collaborative DL systems. 
In the scenario of collaborative learning, all the uploaded data from different users can be trained collaboratively to generate a global model for all users.
However, if an attacker uploads a set of malicious masked data to tamper with the training data, the generated global model will be injected with Trojans.
Following this background, \textit{Auror} firstly analyzes the uploaded masked data in terms of different features and divides them into two groups according to each feature. 
Secondly, once the distance between the two feature groups is larger than the threshold, the corresponding feature can be regarded as the indicative feature, and the group with fewer elements is set as the suspicious group.
Finally, all users are monitored, the user will be regarded as malicious if the masked data uploaded from him triggers the suspicious group more than the set threshold.

Liu \textit{et al.} \cite{Neuron-Trojans-Liu} attempted to detect input samples that diverge from the legitimate distribution of data.
Support Vector Machines (SVMs) \cite{SVM-Hearst} and Decision Trees (DTs) are used as anomaly detection methods.
To tackle the problem that defenders do not know the distribution of illegal data to train SVMs or DTs, the authors propose to set the number of classifiers the same as the number of data classes. 
Specifically, in the classifier training process for the $i$-th class, the data labeled $i$ is taken as positive, and negative otherwise.
Once a benign input sample is input to all the classifiers, it should be classified by one of these classifiers to be positive. 
If no classifier marks the input sample as positive, it is regarded to be abnormal or illegal.

Baracaldo \textit{et al.} \cite{Mitigating-Poisoning-Baracaldo} proposed to use the context information of the original and transformed data in the training set to identify poisoned data. 
The authors segmented the untrusted data into groups where the probability of poisoning is highly correlated between the samples in each group.
Once the training data has been segmented appropriately, the data in each segment is evaluated by comparing the performance of the classifier trained with or without that group. 
The evaluation indicates the detection results, i.e. if the gap in classifier performance between with and without the segmented data group is larger than the threshold, this data group is potentially poisoned. 

Tang \textit{et al} \cite{SCAn-Tang} proposed a Trojan detection method called Statistical Contamination Analyzer (SCAn) based on the statistical properties of the representations produced by an infected model.
A DNN model can generate a representation, which can be decomposed into two parts, including the object's identity and the variation randomly drawn from a distribution.
For example, in the face recognition scenario, the facial features (such as the eyes' color) are related to personal identity, while the posture and expression of one's face are variations. 
In the presence of a Trojan attack, the `Trojan' images change the identity vector and the variation distribution for the target class, rendering them inconsistent with those of other classes. 
Therefore, Trojan detection can be implemented by analyzing the difference between classes. 
However, SCAn requires a set of clean data for the Trojan analysis, which is not feasible in some scenarios. 
Additionally, it relies on the presence of attack images with triggers to identify the potential target class, which still requires prior knowledge of the Trojan attack and the trigger patterns.

\subsubsection{Model-distribution-based Detection}
\ \\

\uline{Model-distribution-based detection is a method that explores the statistical features of model weights and activation feature maps.
This kind of detection method requires the candidate model to be a white box.}

Geigel \textit{et al.} \cite{NN-Trojan-Geigel} used Intrusion Detection Signatures (IDS) to detect Trojans hidden in DNN models, which can be realized by identifying anomalous weights of neural networks. 
This work is based on the assumption that the attacked neural networks have obviously different weight information compared to the original models in the running phase. 
Given a clean DNN model with weights $W$ and a candidate model with weights $W'$ for the same task, we can calculate the offset $p$ between the two models.
With the prior knowledge of Trojan attacks, the offset $p$ can be converged and used as the signature for Trojan detection. 
Once the calculated offset $p$ exists in the offset pool of Trojan attacks, the candidate model is regarded to be attacked.
However, collecting prior knowledge of DNN models is a great challenge for ordinary users.
The detection method based on IDS can only detect the existing Trojan attacks in the library, therefore, we need to make signatures for each attack. 
Additionally, the weights of randomly initialized models may vary greatly after training, although their performance is the same for a specific task.
Thus it is also a great challenge for the summary of prior knowledge. 

To reduce the collecting cost, Ji \textit{et al.} \cite{Programmable-Ji} performed direct statistics on the distribution of neural network weights. 
The author assumed that the weight distribution of the Trojan model is different from that of the benign model.
However, the experimental results show that there is no significant statistical difference between the weights of the benign model and that of the attacked model. 
Therefore, the authors claimed that the static analysis of the model weights can not directly distinguish whether the model is attacked or not. 
Although the proposed weight distribution method is not applicable, it gives us a useful result that analyzing the distribution of the model from a global or statistical perspective can reduce the calculation cost.


To tackle the above problem brought by directly performing statistical analysis, Chan \textit{et al.} \cite{Poison-Cure-Chan} proposed to get the weight distribution of DNN models by calculating the gradient of the loss function to the input sample. 
The authors claimed that the weights connected to these `poison' neurons are much larger than those of other neurons.
According to this assumption, the pixels of the poison pattern have large gradients compared to other pixels. Therefore, if significantly higher gradients are found in an image after calculating the gradient for all the input images, the corresponding input sample is considered to contain the poisoning pattern.

\begin{figure}[!t]
\centering
    \subfigure[]{
    \includegraphics[width=0.2\textwidth]{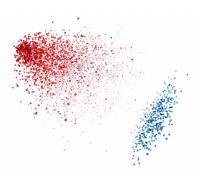}
    }%
		\ \ \ \ \ \
    \subfigure[]{
    \includegraphics[width=0.2\textwidth]{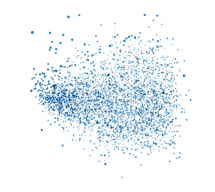}
    }
\caption{Projecting the activation of the last hidden layer onto the first 3 principal components.
(a) poisoned activation; 
(b) unpoisoned activation \cite{Detecting-Backdoor-Chen}.}
\label{fg_Activation-cluster}
\end{figure}




Different from the above methods via performing statistics of model weights, Chen \textit{et al.} \cite{Detecting-Backdoor-Chen} proposed to cluster and analyze the activation values of the last layer to detect Trojans.
It is based on the assumption that the mechanisms of feature extraction of the model are different for the data with and without Trojans.
Given a candidate model, the training data is used to query the network and retain the activation of the last layer.
Once the activation values corresponding to each training sample are  
obtained, they are segmented according to their labels, and each segment is clustered separately. 
To cluster the activation values, the authors reshaped the activation values into two-dimensional vectors by using independent component analysis \cite{ICA}.
As shown in Fig. \ref{fg_Activation-cluster}, the authors used k-means ($k = 2$) to distinguish poisoned activation and benign activation.
Additionally, to locate the cluster corresponding to the poisoned data, the authors introduced three methods, including exclusionary reclassification, relative size comparison, and silhouette score.

Xiang \textit{et al.} \cite{CI-cluster-Xiang} proposed another clustering method to detect Trojans. 
The authors clustered the feature vectors of the penultimate layer and analyzed the distribution of these clusters. 
Different from the method proposed in \cite{Detecting-Backdoor-Chen}, this paper mainly discusses how to distinguish the poisoned clusters in the cluster set, which means that there are possibly multiple clusters for both clean data and trigger patterns, instead of only two clusters. Also, Ji \textit{et al.} \cite{Programmable-Ji} analyzed the distribution of activation values for Trojan detection.
The authors assumed that the activation values of the clean model and the Trojan model are not the same, this is because the activation distribution is highly dependent on the input data and model weights.
However, in a normal training process, two independently trained models may also behave differently in terms of activation distribution, because of the randomly initialized weights and parameters, which means that directly calculating the activation distribution of a model is of limited help in detecting Trojans.

Comparing the two methods proposed in \cite{Detecting-Backdoor-Chen} and \cite{Programmable-Ji}, the latter only makes simple probability distribution statistics on the activation of the model, while the former counts and clusters the activation of the last layer, and then analyzes the clustered results to distinguish the Trojan model.
The reason for this difference between the two methods is discussed by \cite{Programmable-Ji} that the direct analysis on model activation is easily affected by different inputs.
Ma \textit{et al}. \cite{23-detection} proposed `Beatrix', which also performs Trojan detection based on model activation values. However, unlike \cite{Detecting-Backdoor-Chen} and \cite{Programmable-Ji}, `Beatrix' uses Gram Matrices to statistically analyze the model activation values and combine them with input samples for Trojan detection. Specifically, `Beatrix' uses Gram Matrices to capture the correlation of the features of the input samples while capturing the appropriate higher-order information of the representation. By learning the normal samples of the activation pattern of the class conditional statistics, `Beatrix' can capture the anomalies in the activation pattern to identify the poisoned samples. Note that `Beatrix' has a feature that allows it to identify dynamic Trojan attacks (triggers vary from sample to sample) more effectively.

Huang \textit{et al.} \cite{NeuronInspect-Huang} proposed a Trojan detection framework based on output interpretation. 
First, the salience heatmap of the output layer is generated, while the heatmaps generated by the clean model and the Trojan model have different features. 
This is because the Trojan model focuses more on the trigger features in the image than the object features. 
Then, an outlier detector is constructed with three extracted features from the above heatmap, and the median absolute value \cite{Detecting-outliers-Leys} is set as the calculating mechanism. 
Once the normalized anomaly index of a given target label is greater than the threshold, it is considered to be the target class for Trojan attacks. 

\subsubsection{Model-structure-based Detection}
\ \\

\uline{Model-structure-based detection aims to extract abnormal information from the DNN model structure.
It is based on the assumption that Trojans can be mapped into the structure of the attacked model.
This kind of detection method requires the candidate model to be a white box.}

Geigel \textit{et al.} \cite{NN-Trojan-Geigel} proposed a detection method based on code auditing including static analysis and dynamic analysis. 
For static analysis, a proper neural network model has almost no redundant structure, if the complexity of a candidate model exceeds the need for the task, the model is considered to be attacked. 
For dynamic analysis, a detection system is designed as a `data-label' matching program. 
Once the outputs of the candidate model and the detection systems conflict with each other, the alarm of the detection system will be raised, such that the candidate model is regarded to be attacked. 

Differently, Jiang \textit{et al}. \cite{critical-path-Wei} \cite{critical-path-2} proposed a critical-neuron-based Trojan detection method. 
The authors aimed to construct a connection between the model structure and the input samples.
To be specific, the authors claimed that the target class corresponds to images with differently based objects because the poisoned data is usually generated by adding trigger patterns to various classes. 
This will cause the critical neurons corresponding to input samples of the target class to be different.
Based on the above assumption, the critical neurons corresponding to all classes are counted.
The similarity of the critical neurons corresponding to the input samples of the target class is much lower than that of the normal classes. Furthermore, Jiang \textit{et al} also proposed a critical-path-based Trojan detection method that does not rely on any clean dataset. Specifically, they first abstracted a DNN model to a set of critical paths and then built anomaly indices to identify Trojans hidden in the DNN model by calculating the distance and anomaly rate of the critical paths.

\subsubsection{Output-distribution-based Detection}
\ \\

\uline{Output-distribution-based detection is a method to detect Trojans via analyzing the prediction result distribution of a DNN model.
It is based on the observation that the output of a Trojan model generally tends to a certain class. 
This method can work on both black-box and white-box candidate models.}

The authors of \cite{Trojaning-attack-Liu} claimed that the Trojan attacks attempt to mislead the prediction results to the target class. 
Once the input samples are poisoned with the trigger, the model is more likely to output the target class. 
Therefore, a feasible defense is to check the distribution of the prediction results. 
Similarly, Gao \textit{et al.} \cite{STRIP-Gao} also pointed out that the Trojan model would `prefer' the target class. 
Differently, the authors analyzed the confidence of each class and found that the distribution of prediction results of clean data and poisoned data varied much for the Trojan model. 





Yang \textit{et al.} \cite{Generative-Poisoning-Attack-Yang} proposed a method based on the loss value of the model prediction to detect Trojans.
The threat model of this method is the traditional poisoning attack to reduce the performance of DNN models by maximizing the loss term.
Specifically, once the data is input into the candidate model, the loss of the target model is recorded.
If the loss exceeds the loss threshold, a warning is made.
Then, if the number of warnings exceeds the number threshold, an accuracy check is conducted to check if a poisoning attack is in progress.
The proposed method is based on the observation that poisoned input-label pairs have a larger loss value than normal input-label pairs.

Tran \textit{et al.} \cite{Spectral-Signatures-Tran} proposed a concept of `spectral signatures' for Trojan detection.
The method is based on the assumption that the learned representation of the classifier enlarges the critical signal for the classification task, which is defined as the `spectral signatures' of the model. 
Since the Trojan attacks need to change the label of training samples, such that the strong signal of Trojans can also be included in the model. 
Based on the above assumption, the authors suggested finding the hidden `spectral signatures' to detect the potential Trojans. 

\subsubsection{Other Detection Methods}
\ \\

\uline{Most of the above methods are designed to detect Trojans in terms of statistics or distribution. 
Several special detection methods apart from the above detection methods are presented in this subsection.}

Dumford \textit{et al.} \cite{Weight-Perturbations-Dumford} proposed to detect Trojan attacks by periodically calculating the hash function for the model file.
If the hash value is different from the original hash, the model has been changed and may be malicious.
However, this detection method is not entirely reliable because some legal operations can also result in model file changes, such as small random weight perturbations during test \cite{cleverhans-Goodfellow, Deep-learning-Goodfellow} and dropout during test \cite{Dropout-Bayesian-Gal}. 

Chen \textit{et al.} \cite{DeepInspect-Chen} proposed to detect NN Trojans for black-box models.
The defender first applies model inversion on the candidate DNN to recover a substitution training dataset covering all output classes. 
Then, possible trigger patterns belonging to different attack targets are reconstructed by leveraging a generative model and used for Trojan attacks.
Finally, the proposed method formulates Trojan detection as an anomaly detection problem. 
The perturbation statistics in all classes are collected and an outlier indicates the existence of the Trojans.
Differently, Guo \textit{et al.} \cite{TABOR-Guo} formalized the Trojan detection task as a non-convex optimization problem.
Specifically, for a clean model, the local optimum indicates a sub-optimal result. 
In contrast, for a Trojan model, the local optimum may represent the trigger intentionally-injected trigger. 
Therefore, the local optimum can be designed as a key for Trojan detection based on the above assumption.



\begin{figure}[ht]
    \centering
    \includegraphics[width=0.44\textwidth]{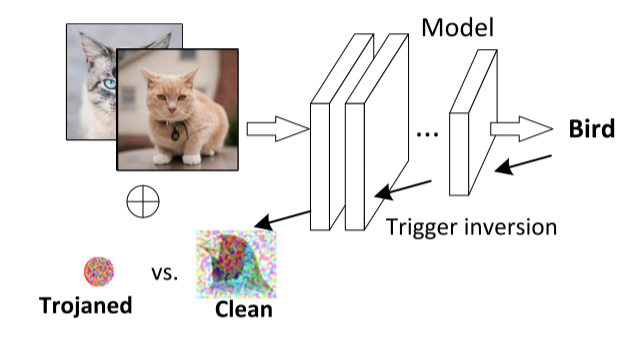}
    \caption{Trojan Detection by trigger inversion.}
    \label{fg_trigger_inversion}
\end{figure}


Additionally, \textit{trigger inversion} is also an effective Trojan detection technique, as shown in Fig. \ref{fg_trigger_inversion}. Neural cleanse (NC) \cite{Neural-Cleanse-Wang}, Artificial Brain Stimulation (ABS) \cite{ABS}, K-Arm \cite{K-arm}, and \cite{22-detection-SFD} all use optimization to invert triggers and determine whether the model is attacked. 
Among them, Ref. \cite{22-detection-SFD} combines trigger inversion with counter-factual causality \cite{counter-factual} to propose a key technique, Symmetric Feature Differencing, a new kind of differential analysis. 
The detection process is that the adversary first obtains a model and a small set of clean samples, then uses trigger inversion methods to generate a specific trigger pattern (e.g., flipping `cat' to `bird').
Next, the victim sample (`cat') and the victim sample with the inverted trigger added are entered into the model. The difference in features in the selected layers of the model is provisionally noted as inverted trigger features (FD1).
Then the difference between the selected layer features of the clean victim sample (`cat") and the clean target class sample (`bird') is noted as the true feature difference (FD2). 
Finally, FD1 and FD2 are compared, and if they have no similarity, the model is considered attacked.

Wang \textit{et al.} \cite{Neural-Cleanse-Wang} proposed to detect Trojans based on the assumption that a Trojan model requires much smaller modifications to cause misclassification into the target class than into other normal classes. 
Therefore, after iterating through all classes, the Trojan can be detected by checking if any class requires a significantly smaller amount of modification to achieve misclassification. 
This method assumes that the Trojan exists in a single target class. 
This assumption, however, fails in an `all-to-all' Trojan attack \cite{BadNets-Gu} where the Trojan exists in all classes. 
To detect the small perturbation, Xiang \textit{et al.} \cite{Optimized-Perturbations-Xiang} proposed an imperceptible Trojan detection method based on perturbations optimization, which is derived from the Test Time Evasion (TTE) attacks \cite{TTS-Szegedy}.
By adding a small, customized perturbation, TTE attacks can make the trained classifier change the decision from one class to another class. 
However, using common perturbations for all images usually requires a large perturbation norm. 
The authors proposed a perturbation optimization algorithm to generate a small perturbation, which enables most of the images embedded with trigger patterns to be misclassified.
Therefore, if this small perturbation can be found, it indicates that DNNs are victims of Trojan attacks.


To build a universal detection method, Xu \textit{et al}. \cite{meta-detection-Xu} proposed a Trojan detection method using meta neural analysis. 
The authors trained a meta-classifier (such as a machine learning model) to detect whether the candidate model is attacked or benign. 
A shadow model is also trained on the same task as the candidate model and requires a small set of clean data. 
Then, the meta-classifier is trained with the outputs from the two models (the candidate model and the shadow model), once the outputs are different, the corresponding input data is labeled as the poisoned data, and the candidate model is regarded as a Trojan model.

\subsection{Comparisons of Trojan Detection Methods}
\label{subsec_detectionAnalysis}

In this section, we review all the Trojan detection methods mentioned above, while the corresponding references are summarized, and extensive summaries and comparisons are presented.
We analyze and summarize each detection method in terms of the black/white box, run-time detection, computing cost, access to Trojan training data, and detection capability, the results are shown in Table \ref{tb_detection_comparison}.
Taking the critical-neuron-based detection method in model-structure-based mechanism as an example, this kind of method requires a white-box DNN model.
It can work in the deployment phase instead of only the training phase.
Applying this method requires moderate computing costs compared to other detection methods, but this method does not require raw training data.
Finally, this method has a 100\% capability for detecting NN Trojans hidden in DNN models. 




\begin{table}[t]
\centering
\footnotesize
\caption{Comparison between different detection methods}
\begin{tabular}{@{}clccccl@{}}
\toprule
Detection Methods                   & \multicolumn{1}{c}{Ref.}            & \begin{tabular}[c]{@{}c@{}}Black/White \\ box\end{tabular} & Runtime & \begin{tabular}[c]{@{}c@{}}Computing\\ Cost\end{tabular} & \begin{tabular}[c]{@{}c@{}}Access to\\ Trojan data\end{tabular} & \multicolumn{1}{c}{\begin{tabular}[c]{@{}c@{}}Detection\\ Capability\end{tabular}} \\ \midrule
\multirow{3}{*}{Data-distribution}    & \cite{AUROR-Shen}                       & Black-box                                                  & No      & Low                                                      & No                                                              & nearly 100\%                                                                       \\
                                           & \cite{Neuron-Trojans-Liu}               & Black-box                                                  & No      & High                                                     & No                                                              & 72.6\% SVM 99\% DT                                                                 \\
                                           & \cite{Mitigating-Poisoning-Baracaldo}   & Black-box                                                  & No      & Moderate                                                 & Yes                                                             & about 92\%                                                                         \\ \hline
\multirow{6}{*}{Model-distribution}  & \cite{NN-Trojan-Geigel}                 & White-box                                                  & No      & Low                                                      & No                                                              & not mentioned                                                                      \\
                                           & \cite{Programmable-Ji}                  & White-box                                                  & No      & Low                                                      & Yes/No                                                          & very poor                                                                          \\
                                           & \cite{Poison-Cure-Chan}                 & White-box                                                  & Yes     & High                                                     & Yes                                                             & about 99\%                                                                         \\
                                           & \cite{Detecting-Backdoor-Chen}          & White-box                                                  & No      & Moderate                                                 & Yes                                                             & F1 score nearly 100\%                                                              \\
                                           & \cite{CI-cluster-Xiang}                 & White-box                                                  & No      & Moderate                                                 & Yes                                                             & more than 93\%                                                                     \\
                                           & \cite{NeuronInspect-Huang}              & White-box                                                  & Yes     & Low                                                      & Yes                                                             & nearly 100\%                                                                       \\
                                           & \cite{23-detection}              & White-box                                                  & Yes/No     & High                                                      & Yes                                                             & F1 score over 90\%
                                           \\
                                           & \cite{22-detection-SFD}              & White-box                                                  & No     & High                                                      & Yes                                                             & around 85\%
                                           \\
                                           \hline
\multirow{2}{*}{Model-structure}     & \cite{NN-Trojan-Geigel}                 & White-box                                                  & No      & Low                                                      & No                                                              & not mentioned                                                                      \\
                                           & \cite{critical-path-Wei}                & White-box                                                  & Yes     & Moderate                                                 & No                                                              & 100\%                                                                              \\ \hline
\multirow{4}{*}{Result-distribution} & \cite{Trojaning-attack-Liu}             & Black-box                                                  & No      & Moderate                                                 & Yes                                                             & not mentioned                                                                      \\
                                           & \cite{STRIP-Gao}                        & Black-box                                                  & Yes     & Low                                                      & No                                                              & 0.46\% FAR and 1\% FRR                                                             \\
                                           & \cite{Generative-Poisoning-Attack-Yang} & Black-box                                                  & No      & Low                                                      & Yes                                                             & nearly 100\%                                                                       \\
                                           & \cite{Spectral-Signatures-Tran}         & Black-box                                                  & No      & Moderate                                                 & Yes                                                             & about 90\%                                                                         \\ \hline
\multirow{5}{*}{Other detection}   & \cite{Weight-Perturbations-Dumford}     & White-box                                                  & No      & Moderate                                                 & No                                                              & not mentioned                                                                      \\
                                           & \cite{DeepInspect-Chen}                 & Black-box                                                  & No      & High                                                     & Yes                                                             & 1.56\% FRR                                                                         \\
                                           & \cite{TABOR-Guo}                        & White-box                                                  & No      & Low                                                      & Yes                                                             & 60\% F1 and 55\% recall                                                            \\
                                           & \cite{Neural-Cleanse-Wang}              & Black-box                                                  & No      & High                                                     & No                                                              & not mentioned                                                                      \\
                                           & \cite{Optimized-Perturbations-Xiang}    & Black-box                                                  & No      & Moderate                                                 & No                                                              & about 93\%                                                                         \\ \bottomrule
\end{tabular}
\label{tb_detection_comparison}
\end{table}

\begin{table}[!b]
\centering
\footnotesize
\caption{Analysis of advantages and disadvantages of different Trojan detection methods}
\begin{tabular}{@{}cll@{}}
\toprule
Detection Methods & \multicolumn{1}{c}{Advantages}                                                         & \multicolumn{1}{c}{Disadvantages}                                                                                              \\ \midrule
Data-distribution &  Easy to implement.                                                                 & \begin{tabular}[c]{@{}l@{}}Reliance of the dataset, \\ requirement of prior knowledge of data distribution.\end{tabular}        \\
Model-distribution       & Intuitive.                                                                              & \begin{tabular}[c]{@{}l@{}}Requirement of weights, \\ high computing cost,\\ prior knowledge of Trojan signatures.\end{tabular} \\
Model-structure          & \begin{tabular}[c]{@{}l@{}}High detection performance,\\ interpretability.\end{tabular} & Requirement of white-box model.                                                                                                 \\
Result-distribution      & \begin{tabular}[c]{@{}l@{}}Easy to implement, \\ intuitive.\end{tabular}            & Requirement of prior distribution knowledge.                                                                                   \\ \bottomrule
\end{tabular}
\label{tb_detection_advantages}
\end{table}

Table \ref{tb_detection_advantages} shows the advantages and disadvantages of the five kinds of detection methods against NN Trojans.
For example, the detection methods based on the model-structure mechanism have a high detection performance.
Especially, the critical-neuron-based detection method can use the interpretability of DNN models to detect the anomaly behavior of the model caused by Trojans. 
However, these methods require the candidate models to be white-box, which is the biggest limitation.

\subsection{Research Challenges on Trojan Detection}
\label{subsec-challenge-detection}

In this section, we give the possible development directions and challenges of neural network Trojan detection methods.
The existing detection methods are mainly constructed based on macro observation, such as the data distribution analysis. 
We can start from two aspects to improve the efficiency and accuracy of detection methods. 
One is model-based mathematical analysis, and the other is to build more targeted detection methods for each kind of attack.
The limitations of detection methods for NN Trojans are summarized as follows:

\begin{itemize}
\item The current detection methods are mostly passive.
\item There are fewer detection methods against hardware Trojans and research on hardware Trojan attacks is lacking.
\item There are no mature systematic detection or restoration platforms. 
Decentralized detection methods are aimless and inefficient.
\end{itemize}

For the detection methods against Trojan attacks, we also conduct an analysis of the possible development directions:

\begin{itemize}
\item Constructing and completing the formal detection methods, from the perspective of mathematical analysis.
\item Since most of the application scenarios are black-box, efficient detection methods for black-box models are necessary. 
\item Constructing the detection platform or system to reduce the cost of detection.
\item Designing more detection methods against hardware Trojans, e.g., power consumption detection and hardware application of anomaly detection.
\item Constructing a complete set of application processes of DNNs for production, supervision, anti-stealing, and detection.
\end{itemize}

In summary, building a feature bank for various Trojans, constructing integrated platforms, etc., are the feasible developing directions for Trojan detection in the future. 

\section{Defense Methods against Neural Network Trojans}
\label{sec_defense}

In this section, we discuss and compare the defense methods for DNN models against Trojan attacks. They can be based on Trojan detection results to remove Trojans from the DNN model and mitigate the impact caused by Trojans preeminently. As in the previous two sections, this section elaborates on four aspects, namely, principles (Sec. \ref{subsec_defense_principle}), classification and corresponding specific defense methods (Sec. \ref{subsec_list_of_Trojan_defense}), comparisons (Sec. \ref{subsec_defenseAnalysis}), and potential challenges (Sec. \ref{subsec-challenge-defense}).

\subsection{Principle of Trojan Defense}
\label{subsec_defense_principle}

\begin{figure}[b]
\centering
\includegraphics[scale=0.42]{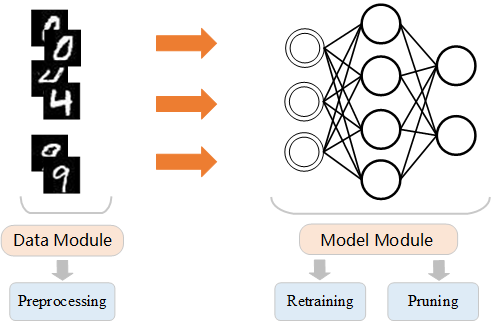}
\caption{Defense from different perspectives.}
\label{fg_defense_principle}
\end{figure}

Compared with Trojan detection methods, Trojan defense can be conducted from two perspectives, including input data processing and model processing, as shown in Fig. \ref{fg_defense_principle}.
First, most of the Trojans require poisoned data as the trigger to start the attack.
Therefore, it makes sense to pre-process the input samples to defend against the Trojans, as the trigger probability can be reduced by filtering suspicious input samples.
Second, as mentioned in Sec. \ref{subsec_list_of_Trojan_detection}, most of the Trojan attacks correspond to a specific structure or a sub-network of the model. 
These Trojan attacks can be divided into two parts, including unspecified injected Trojans and specified injected Trojans.
All the data-driven Trojans and most of the model-based Trojans are unspecified injected Trojans, meaning that there is no need to specify the sub-network corresponding to the Trojan.
In contrast, part of the model-based Trojans and all neuron Trojans are specified injected Trojans. 
It is intuitive to design a method to manipulate the model for Trojan defense, such as retraining and pruning operations on DNN models.

\subsection{State-of-the-Arts Defense Methods}
\label{subsec_list_of_Trojan_defense}

\begin{figure}[t]
\centering
\includegraphics[scale=0.4]{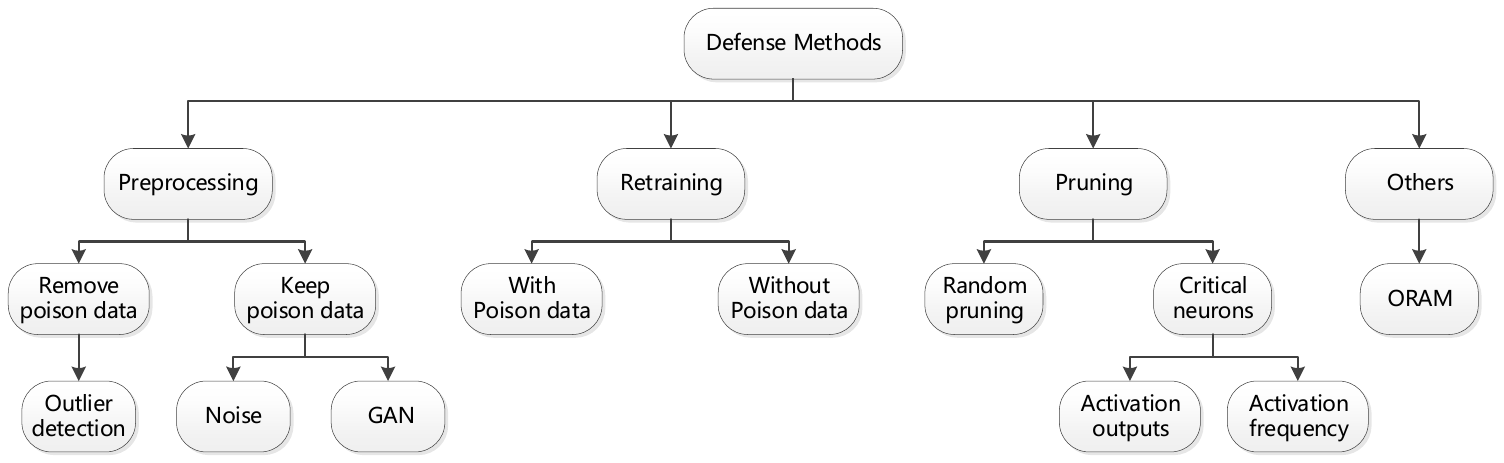}
\caption{Classification of Trojan defense methods.}
\label{fg_Trojan_defense_classification}
\end{figure}

In this section, we review several efforts on Trojan defense methods. We tend to divide the defense methods into four categories, including \textit{pre-processing, retraining, pruning,} and \textit{other defense} methods.
As shown in Fig. \ref{fg_Trojan_defense_classification}, each kind of Trojan defense method can also be subdivided from the perspective of implementation methods. 
Taking the pruning-based Trojan defense method as an example, it can be conducted from two aspects, i.e., random pruning and critical neuron pruning (fine-pruning).
The random pruning method can be regarded as a universal defense method for different NN Trojans, while the fine-pruning method is more precise and can be implemented in terms of activation output statistics and activation frequency of neurons.
The following content introduces and discusses the implementation of each defense strategy through different aspects.

\subsubsection{Pre-processing-based Defense}
\ \\

\uline{Pre-processing-based defense is a method that filters the potential poisoned data during the deployment phase.
This kind of defense method requires only the filtering mechanism and the input samples.}

\begin{figure}[t]
\centering
\begin{minipage}[t]{0.46\textwidth}
    \centering
    \raisebox{0.2\height}{
        \includegraphics[width=6.6cm]{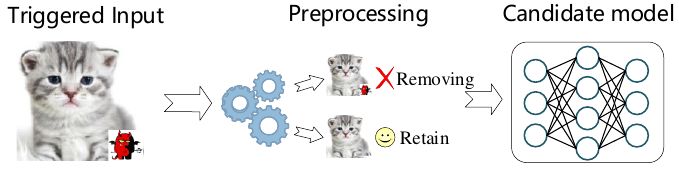}}
    \caption{Pre-processing the input data for defense.}
    \label{fg_preprocessing}
\end{minipage}
\ \ \ \ \ \
\begin{minipage}[t]{0.5\textwidth}
\centering
\includegraphics[width=6.8cm]{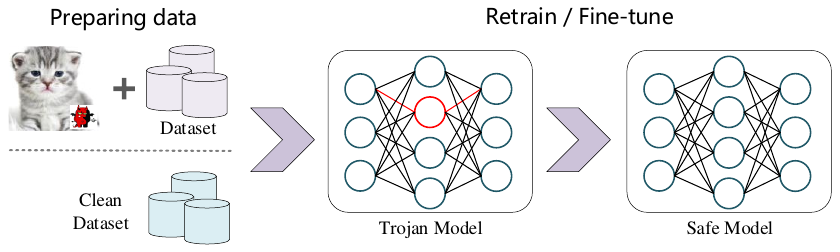}
\caption{Retraining with/without poisoned data.}
\label{retraining}
\end{minipage}
\end{figure}

Most Trojan attacks need some specific inputs to trigger the Trojans.
Once the trigger input is fed into the Trojan model, the Trojan model can output malicious results.
To tackle this problem, we can discard or repair the input samples in the run-time system, as shown in Fig. \ref{fg_preprocessing}.


Liu \textit{et al.}\cite{Neuron-Trojans-Liu} proposed to use the auto-encoder \cite{autoencoder-Hinton} to pre-process input samples.
The purpose of pre-processing is to prevent poisoned input from triggering the Trojan without affecting the accuracy of normal data.
Note that only clean data is used to train the auto-encoder.
In the deployment phase, if the input samples are in the distribution of the clean data, the output of the auto-encoder should be close to the input. 
In this case, the candidate model can classify the reconstructed image as accurately as the original input. 
However, if the input sample is out of the distribution, the auto-encoder will also reconstruct the input according to the trained distribution. 
The reconstruction causes a large deviation between the original poisoned inputs and auto-encoder outputs, which may disable the Trojan attack.

Yao \textit{et al.} \cite{Latent-Backdoor-Yao} proposed an assumption that adding small perturbations to the input samples can mitigate the Trojan attacks.
The intuition behind this method is that general Trojan attacks need well-designed trigger patterns, and these patterns are sensitive to pixel-level perturbation. 
Therefore, the authors applied the Gaussian filter to remove the noise perturbation from the input image, and then the processed images were put into the model. 
However, although adding perturbation can reduce the attack success rate of Trojans, it also reduces the model accuracy on benign data.

Additionally, Doan \textit{et al.} \cite{Februus-Doan} used a Generative Adversarial Network (GAN) to repair the input samples that may be poisoned. 
First, they use Gradient-weighted Class Activation Mapping (GradCAM) \cite{Grad-CAM-Selvaraju} to generate a heatmap to illustrate the important areas in the input samples which is the key to the learned features.
After obtaining the key area, it is directly removed from the original image. 
Then the trained GAN model is utilized to re-fill the images after removing areas. 
Note that the data for training the GAN model has the same distribution of clean data for training the Trojan model. 

Different from the defense scenario of the deployment phase, Steinhardt \textit{et al}. \cite{Certified-Defenses-Steinhardt} proposed that it is important to defend against Trojan attacks in the training phase. 
Specifically, the whole dataset will be checked, then poisoning points (outliers) will be removed and a margin-based loss on the remaining data will be minimized. 
Note that the threat model in this paper is about traditional poisoning attacks which aim to reduce the classification performance of DNN models.

\subsubsection{Retraining-based Defense}
\ \\

\uline{Retraining-based defense is a method that uses new data to retrain or fine-tune the potentially attacked model.
This kind of defense method is generally applied before the deployment phase of DNN models.}

A feasible way for Trojan defense is to regard the model as a black box and use new data to fine-tune it.
The intuition behind this operation is that retraining can change the distribution and the functions of the weights in the candidate model.
As shown in Fig. \ref{retraining}, for a black-box model without knowing the exact meaning of each neuron or weight, retraining or fine-tuning can be utilized to adjust the weight distribution or local value to reduce the risk of unreliable models.


Liu \textit{et al.} \cite{Neuron-Trojans-Liu} claimed that retraining may reduce the risk of potentially attacked models.
The Trojans contained in the malicious weights may be covered during the retraining process since the Trojans are sensitive to the malicious weights.
Additionally, although the strategy of retraining is similar to that of training from scratch, much fewer training samples are used in retraining.
It results in faster convergence, and the workload of retraining is much less than that of a training model from scratch.

Baracaldo \textit{et al.} \cite{Mitigating-Poisoning-Baracaldo} re-applied the separated poisoned data from the dataset for retraining.
Such scenarios are regarded as online learning or regular training.
However, if all the poisoned data is directly utilized for retraining, it may cause poor model accuracy on the benign data, even if the fault tolerance is slightly improved.
Therefore, the authors proposed to first remove the poisoned data that is different from the clean data, and then the remaining data is retrained online, which is possible to improve the robustness of the candidate model while maintaining high accuracy.
Similarly, Chen \textit{et al.} \cite{Detecting-Backdoor-Chen} also proposed that retraining is a feasible way to eliminate Trojans while improving the robustness of the model.
The authors claimed that removing the poisoned data and retraining the model from scratch with the remaining clean data is time-consuming. 
A faster method is to re-label the poisoned data with the correct class and then re-train the model with them. 

For Trojan attacks using data poisoning, Chan \textit{et al.} \cite{Poison-Cure-Chan} proposed a gradient-based method to get poison signals and a retraining-based defense method.
The authors claimed that the similarity between the gradients on poisoned inputs and poisoning signals is higher than that of clean inputs. 
Due to the lack of trigger patterns in clean samples, most of the Trojan neurons are not activated during the inference process, resulting in almost no poison signals in the gradients on input samples. 
To eliminate the negative effects of Trojans and improve the robustness of the candidate model, poison signals are added to all the classes. 
Then retraining strategy is applied to reduce the association between the trigger pattern and the target class.

Similar to \cite{Poison-Cure-Chan}, Chen \textit{et al}. \cite{22-defense-sensitivity} found a significant difference in the sensitivity of poisoned samples and clean samples to transformations.
Therefore, they devised a simple sensitivity metric, called Feature Consistency towards Transformations (FCT), to distinguish poisoned samples from clean samples in the poisoned training data. Based on FCT, they proposed two effective Trojan defense methods. The first approach uses a two-stage security training module (called Distinguished and Secure Training) to train the security model from scratch. Specifically, the first sample differentiation module divides the entire training set into clean, poisoned, and uncertain samples based on the FCT metric. The second two-stage secure training module learns the feature extractor by semi-supervised contrastive learning and then the classifier by minimizing the mixed cross-entropy loss. The second approach then uses a Trojan removal module to remove Trojans from the backdoor model. Specifically, the Trojan removal module alternately unlearns the distinguished poisoned samples and re-learns the distinguished clean samples, thus removing the effect of poisoned samples from the Trojan model.

Based on the assumption that NN Trojans are hidden in weights, Ji \textit{et al.}\cite{Programmable-Ji} proposed to fine-tune models for Trojan defense. 
Compared to the above retaining-based methods \cite{Mitigating-Poisoning-Baracaldo} and \cite{Detecting-Backdoor-Chen} requiring the part of clean data after separating poisoned data, this method only needs a small-scale clean dataset for the fine-tuning of the convolutional layers instead of the whole model. 
The authors claimed that fine-tuning with only the clean dataset can also reduce the risks brought by Trojan attacks.

Different from the method of direct retraining, Qiao \textit{et al.} \cite{Generative-Distribution-Qiao} tried to get the Max-Entropy Staircase Approximator (MESA) before retraining to calculate the trigger distribution of the Trojan model.
The distribution knowledge is applied for the model repair. 
MESA integrates a set of sub-models to approximate the unknown trigger distribution, and each sub-model only needs to learn part of the distribution. 
Specifically, a batch of noise is randomly generated as potential triggers and sent to the Trojan model and the sub-models, respectively. 
Then, these two branches calculate the entropy of output and the entropy of triggers. 
The combined entropy losses are then used to update the trigger generators and the sub-models. 
Any class found by MESA that triggers an attack success rate higher than the threshold is considered to be attacked.
For each target class, re-running MESA to get several sub-models. For each sub-model, the Trojan can be eliminated by model retraining.
Finally, evaluate the retrained models to decide which sub-model produces the best defense.

Zhao \textit{et al}. \cite{AI-lancet-Zhao} proposed to first locate the error-inducing neurons that play a key role in generating the erroneous output, and then repair the model via neuron-flip and neuron-fine-tuning.
To be specific, the key areas of the misclassified samples are located first.
The companion sample corresponding to this sample can be generated by removing the located area.
After performing the differential feature analysis between the original sample and the companion sample, the feature leading to the misclassification can be revealed.
The contribution of the neurons to the features is then calculated.
The error-inducing neurons can be located via the progressive-ablation method on these contributions.
Finally, the authors proposed two methods to repair these error-inducing neurons via neuron flipping and neuron fine-tuning.
The former is implemented by reversing the sign of values to break the connection to the erroneous output, which is in no need of any training data.
In contrast, the latter needs extra training samples to fine-tune the selected error-inducing neurons to change their values to break the propagation of Trojans.

\subsubsection{Pruning-based Defense}
\ \\

\uline{Pruning-based defense is a method that aims to prune neurons connected to the Trojans hidden in the candidate model.}

\begin{figure}[t]
\centering
\includegraphics[scale=1.0]{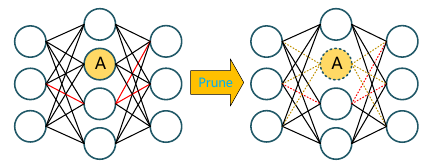}
\caption{Two typical pruning methods for neural networks. 
1) Pruning the neuron (such as neuron `A') and the weights (yellow lines) connected to it.
2) Pruning only the selected weights (red lines).}
\label{pruning}
\end{figure}

A neural network consists of numerous neurons and is connected by weights.
Trojans are usually hidden in neurons or generated by the connection between neurons.
Therefore, as shown in Fig. \ref{pruning}, pruning-based defense methods are intuitive to break the connections or prune malicious neurons to reduce the risks brought by Trojan models.

As mentioned above, retraining and fine-tuning-based methods can reduce the negative effects of Trojans.
However, retraining is costly, especially for large datasets or large-scale networks.
Additionally, Liu \textit{et al.} \cite{Fine-Pruning-Liu} even claimed that the fine-tuning defense is not always workable against Trojan models.
This is because the inference of clean inputs does not rely on the Trojan neurons, and these neurons are dormant during the inference for clean inputs.
Therefore, the fine-tuning operation may not update the weights connected to these Trojan neurons, leaving them unchanged.
Based on the above assumption, Liu \textit{et al.} \cite{Fine-Pruning-Liu} proposed a method called fine-pruning, which aims to combine both pruning and fine-tuning for Trojan defense.
The authors first prune the candidate DNN model, and then fine-tune the pruned network, while pruning can remove Trojan neurons and fine-tuning can recover the accuracy loss on the clean data introduced by pruning.

Wang \textit{et al.} \cite{Neural-Cleanse-Wang} used the neuron sensitivity difference between poisoned data and clean data to select Trojan neurons. 
Both clean data and poisoned data are fed to the candidate model. 
The activation outputs of hidden layers are collected to compare the difference in the activation output of neurons under different input samples. 
When some neurons are sensitive to poisoned inputs and insensitive to clean inputs, they are regarded as Trojan neurons. 
Then, the pruning strategy is applied to stop the Trojan propagation among them by setting the corresponding weights to zero.

To improve the pruning precision, Jiang \textit{et al}. \cite{critical-path-Wei} proposed to locate the Trojan neurons in the model via the critical-neuron-based method.
Different from the method proposed in \cite{Neural-Cleanse-Wang} counting for the activation outputs in layers, the authors counted for the activation frequency of neurons indicated by control gates. 
The control gates are equipped following neurons in each layer, and trained to indicate the sensitivity of neurons on various input samples.
This paper assumes that different neurons in a model are critical to different input pattern features, and the trigger pattern also has a strong connection to some specific neurons in the model.
Therefore, the neurons with a strong connection to trigger patterns are collected as the Trojan neurons, and the pruning strategy is applied to prune these neurons from the Trojan model. 
The experimental results showed that this pruning strategy can eliminate Trojan attacks by only pruning a small number of Trojan neurons while maintaining a high accuracy on clean data without any retraining.

\subsubsection{Other Defense Methods}
\ \\
Defense methods against hardware-based Trojans are mainly discussed in this section.
Most of the hardware-based Trojans are based on neural network accelerators.
Zhao \textit{et al.} \cite{Memory-Attack-Zhao} proposed to inject Trojans into the memory control unit to make the neural network accelerator produce malicious results.
The authors also proposed a defense method against this kind of Trojans via using oblivious RAM (ORAM) \cite{ORAM-Stefanov}.
ORAM is utilized to hide the memory access mode making the Trojan fail to recognize read-and-write operations, thereby preventing attacks that use the read-and-write information of memory.
However, the authors also mentioned that the ORAM strategy significantly increases the number of memory accesses and brings a huge memory overhead.
Therefore, this strategy is not realistic on memory-intensive platforms such as neural network accelerators.

\subsection{Comparisons of Trojan Defense methods}
\label{subsec_defenseAnalysis}
\begin{table}[!b]
\centering
\footnotesize
\caption{Comparison between different defense methods}
\begin{tabular}{@{}clccccl@{}}
\toprule
Defense Methods                 & \multicolumn{1}{c}{Ref.}          & \begin{tabular}[c]{@{}c@{}}Black/White\\ Box\end{tabular} & Runtime & \begin{tabular}[c]{@{}c@{}}Computing\\ Cost\end{tabular} & \begin{tabular}[c]{@{}c@{}}Access to\\ Trojan data\end{tabular} & \multicolumn{1}{c}{\begin{tabular}[c]{@{}c@{}}Defense\\ Capability\end{tabular}} \\ \midrule
\multirow{4}{*}{Pre-processing} & \cite{Neuron-Trojans-Liu}             & Black-box                                                 & Yes     & Moderate                                                 & No                                                              & About 90\%                                                                       \\
                                & \cite{Latent-Backdoor-Yao}            & Black-box                                                 & Yes     & Low                                                      & No                                                              & 80\%                                                                             \\
                                & \cite{Februus-Doan}               & Black-box                                                 & Yes     & High                                                     & No                                                              & Nearly 100\%                                                                     \\
                                & \cite{Certified-Defenses-Steinhardt}  & Black-box                                                 & No      & High                                                     & Yes                                                             & 97\%                                                                             \\ \hline
\multirow{7}{*}{Retraining}     & \cite{Neuron-Trojans-Liu}             & White-box                                                 & No      & Moderate                                                 & No                                                              & 94\%                                                                             \\
                                & \cite{Mitigating-Poisoning-Baracaldo} & White-box                                                 & Yes     & Moderate                                                 & Yes                                                             & TPR 0.938 FPR 0.163                                                              \\
                                & \cite{Detecting-Backdoor-Chen}        & White-box                                                 & No      & Moderate                                                 & Yes                                                             & More than 90\%                                                                   \\
                                & \cite{Neural-Cleanse-Wang}            & White-box                                                 & -       & -                                                        & -                                                               & -                                                                                \\
                                & \cite{Programmable-Ji}                & White-box                                                 & No      & Low                                                      & No                                                              & Average 70\%                                                                     \\
                                & \cite{Generative-Distribution-Qiao}   & White-box                                                 & No      & High                                                     & No                                                              & About 99\%                                                                       \\
                                & \cite{Poison-Cure-Chan}               & White-box                                                 & No      & Moderate                                                 & Yes                                                             & About 97\%                                                                       \\
                                & \cite{22-defense-sensitivity}               & White-box                                                 & No      & Moderate                                                 & Yes                                                             & About 90\%
                                \\
                                \hline
\multirow{3}{*}{Pruning}        & \cite{Fine-Pruning-Liu}               & White-box                                                 & No      & Low                                                      & No                                                              & About 70\%                                                                       \\
                                & \cite{Neural-Cleanse-Wang}            & White-box                                                 & No      & Low                                                      & Yes                                                             & About 90\%                                                                       \\
                                & \cite{critical-path-Wei}              & White-box                                                 & Yes     & Moderate                                                 & No                                                              & More than 95\%                                                                            \\ \hline
Other methods                   & \cite{Memory-Attack-Zhao}             & Black-box                                                 & Yes     & Low                                                      & No                                                              & 100\%                                                                            \\ \bottomrule
\end{tabular}
\label{tb_defense_comparison}
\end{table}
In this section, we review all the Trojan defense methods mentioned above, while the corresponding references are summarized, and extensive summaries and comparisons are presented.
We analyze and summarize each defense method in terms of the black/white box, run-time detection, computing cost, and access to Trojan training data for detection and detection capability, the results are shown in Table \ref{tb_defense_comparison}.
Take the activation-frequency-based defense method in pruning-based mechanism as an example, this kind of method requires to analyze the activation frequency of neurons in the candidate model.
It can work in the deployment phase by collecting run-time input samples.
Applying this method requires moderate computing cost compared to other defense methods since it needs to train the control gate for the activation frequency statistics.
The original training data is not required for this method to inject Trojans.
Finally, this method can eliminate more than 95\% of the potential Trojan attacks. 

\begin{table}[t]
\centering
\footnotesize
\caption{Analysis of advantages and disadvantages of different Trojan defense methods}
\begin{tabular}{@{}cll@{}}
\toprule
Defense Methods & \multicolumn{1}{c}{Advantages}                                                                                                                              & \multicolumn{1}{c}{Disadvantages}                                                                                                                   \\ \midrule
Pre-processing    & \begin{tabular}[c]{@{}l@{}}No need to modify the candidate model,\\ work in the run-time scenario,\\ the completion-based defense is universal.\end{tabular} & \begin{tabular}[c]{@{}l@{}}Samples that evade inspection remain a threat,\\ generative models are costly.\end{tabular}                               \\
Retraining        & \begin{tabular}[c]{@{}l@{}}Universal, \\ a small amount clean data is required.\end{tabular}                                                                  & \begin{tabular}[c]{@{}l@{}}Low effectiveness,\\ non-targeted.\end{tabular}                                                                           \\
Pruning           & \begin{tabular}[c]{@{}l@{}}Intuitive and effective,\\ targeted,\\ accuracy loss is low on clean data.\end{tabular}                                           & \begin{tabular}[c]{@{}l@{}}Hard to locate abnormal neurons,\\ random pruning strategies still need retraining \\ for recovering accuracy.\end{tabular} \\ \bottomrule
\end{tabular}
\label{tb_defense_advantages}
\end{table}

Table \ref{tb_defense_advantages} shows the advantages and disadvantages of these three kinds of defense methods against NN Trojans.
For example, the defense methods based on the pre-processing mechanism do not need to modify the candidate model and can work in the run-time scenario.
This kind of method can be combined with the generative strategy to eliminate the poisoned data and repair the images.
However, once the poisoned sample escaped the processing and still maintains the poisoning function, the candidate model can still get malicious outputs, since the Trojan model is not changed. 

\subsection{Research Challenges on Trojan Defense}
\label{subsec-challenge-defense}

In this section, we give the possible development directions and challenges of neural network Trojan defense methods.
The existing defense methods are mainly constructed based on processing potentially poisoned data or repairing the Trojan model. 
We can start from two aspects to improve the efficiency of defense methods. 
One is to construct a more effective data-checking and processing method, and the other is to build more targeted pruning-based methods.
The limitations of Trojan defense methods for DNN models are as follows:

\begin{itemize}
\item The current noise-based defense method is non-targeted, and may cause accuracy loss on clean data. Additionally, the fine-pruning-based defense methods are limited to model structures, which means they are not universal for various DNN models.
\item GAN-based pre-processing methods are time-consuming to train the generative model, which is rarely universal for different datasets.
\item Retraining the model with malicious data after correcting labels will reduce the accuracy of the model.
\item There are fewer defense methods against hardware Trojans and no mature defense platforms or systems for repairing the Trojan model. 
\end{itemize}

For the defense methods against Trojan attacks, we also analyze the possible development directions:

\begin{itemize}
\item Training a universal generative model with a large-scale dataset to repair the damaged images.
\item Pre-processing is suitable for the scenario of DNN models since it is proper to the application scenarios of black-box DNN models. 
\item Constructing a defense platform or system to reduce the cost of defense.
\item More fine-grained pruning strategy is needed for designing the pruning-based defense methods and designing a universal pruning method for different model structures.
\item Designing more defense methods against hardware Trojans, e.g., removing the potentially attacked circuits.
\end{itemize}

As a conclusion, the precise pruning strategy is proper for white-box applications and the pre-processing strategy is a key direction to mitigate the Trojan attacks in a black-box scenario.

\section{Conclusion}

This paper comprehensively reviews the impact of Trojan attacks against DNN models.
Although DNNs have high accuracy in a variety of classification or regression tasks, 
they have been found to be susceptible to subtle input disturbances, which are able to cause complete output change.
As DNNs become the core of current ML-based and AI-based applications, 
this discovery has inspired many contributions to neural network Trojan attacks and defenses. This paper reviews these contributions, 
focusing on the most influential and interesting works in the literature.
Existing literature shows that current DNNs can not only effectively attack in cyberspace, 
but also effectively attack the physical world.
In summary, with the widespread application of third-party models, NN Trojans will generate greater threats, especially in safety and security-critical applications, which shows the importance of calling attentions to the security issues of DNN-based applications.

\bibliographystyle{ACM-Reference-Format}
\bibliography{sample-base}

\end{document}